\documentclass[conference]{IEEEtran}
\IEEEoverridecommandlockouts
\usepackage[T1]{fontenc} 
\usepackage{amssymb}
\usepackage[hyperfootnotes=false]{hyperref}
\usepackage{cite}
\usepackage{graphicx}
\usepackage{array}
\usepackage{multirow}
\usepackage{amsmath}
\usepackage{stfloats}
\usepackage{graphicx}
\usepackage{subfigure}
\usepackage{tabularx}
\usepackage{booktabs}
\usepackage{epsfig,epsf}
\usepackage{setspace}
\usepackage{bm}
\usepackage{textcomp}
\usepackage{algorithmic}
\usepackage{algorithm}
\usepackage{subfigure}
\usepackage{caption}
\usepackage{graphicx}
\usepackage{setspace}
\usepackage{url}
\usepackage{verbatim}
\usepackage{graphicx}
\usepackage{cite}
\usepackage{amssymb}
\usepackage{color}
\usepackage{xpatch}
\definecolor{shadecolor}{RGB}{0,0,255}
\definecolor{blue}{RGB}{0,0,255}

\newtheorem{proposition}{Proposition}
\newtheorem{corollary}{Corollary}
\newtheorem{remark}{Remark}
\usepackage{soul}
\usepackage{booktabs}
\usepackage{tabularx}
\usepackage{array}
\usepackage[table]{xcolor}
\usepackage{braket}

\newcolumntype{Y}{>{\raggedright\arraybackslash}X}

\definecolor{tableblue}{RGB}{220,230,241}

\makeatletter

\ExplSyntaxOn
\cs_new:Npn \bibColoredItems #1#2
{
	\clist_map_inline:nn {#2} { \cs_new:cpn {bib@colored@##1} {#1} } 
}
\ExplSyntaxOff

\newcommand\bib@setcolor[1]{%
	\ifcsname bib@colored@#1\endcsname
	\expanded{\noexpand\color{\csname bib@colored@#1\endcsname}}%
	\else
	\normalcolor
	\fi
}

\xpretocmd{\@bibitem}{\bib@setcolor{#1}}{}{}
\xpretocmd{\@lbibitem}{\bib@setcolor{#2}}{}{}
\makeatother
\bibColoredItems{black}{peng2026rfparadigmrydbergatomic,Gongtcom2026,yan2026enhancinginstantaneousbandwidthrydberg,ITU2023IMT2030,Heydari2021THz,Sedlacek2012Electrometry,Jing2020Superheterodyne,Meyer2018Digital,Anderson2021AMFM,Hu2023Bandwidth,Meyer2023Multiband,Wu2025LODC,Tao2026OFDM,Wang2026Multicarrier,Chen2026SWM,Cui2026Multiband,Saad2020Vision6G,Tataria2021Wireless6G,Lin2015WidebandReceivers,Sharma2020ReceiverTradeoffs,Mohapatra2007RydbergEIT,Holloway2019PhaseMod,Deb2018RadioFiber,Cai2023PhaseFDM,Yuan2023MillimeterAM,Nowosielski2024QAM,Gong2025WirelessOverview,Gong2025MIMOUplink,Borowka2025OpticalBias,Nan2026FHSS,Berweger2023Tuning,Song2024Continuous,Bohaichuk2022Transient,Yang2024Instantaneous,Tang2025Dynamic,Schlossberger2026Linewidth,Ren2024Dynamic}
\begin{document}
	
\title{{Continuous Wideband Multicarrier Rydberg Atomic Quantum Receiver}}

\author{Qihao Peng,~\IEEEmembership{Member,~IEEE},
         Qu Luo,~\IEEEmembership{Member,~IEEE},\\
          Maged Elkashlan,~\IEEEmembership{Senior Member,~IEEE},
        Cunhua Pan,~\IEEEmembership{Senior Member,~IEEE},
       Pei Xiao,~\IEEEmembership{Senior Member,~IEEE},\\
       George K. Karagiannidis,~\IEEEmembership{Fellow,~IEEE},
       Trung Q. Duong,~\IEEEmembership{Fellow,~IEEE},
        Jiangzhou Wang,~\IEEEmembership{Fellow,~IEEE}.
   \vspace{-1cm}
		\thanks{Q. Peng, Q. Luo, and P. Xiao are affiliated with 5G and 6G Innovation Centre, Institute for Communication Systems (ICS) of the University of Surrey, Guildford, GU2 7XH, UK. (e-mail: \{q.peng, q.u.luo,p.xiao\}@surrey.ac.uk). } \\
         \thanks{Maged Elkashlan is with Queen Mary University of London, United Kingdom. (E-mail: maged.elkashlan@qmul.ac.uk)} \\
  \thanks{C. Pan and J. Wang are with the National Mobile Communications Research Laboratory, Southeast University, Nanjing, China. (e-mail: cpan@seu.edu.cn;j.z.wang@seu.edu.cn).}  \\
        \thanks{George K. Karagiannidis is with
the Department of Electrical and Computer Engineering, Aristotle University of Thessaloniki, Thessaloniki 54124, Greece (e-mail: geokarag@auth.gr).} \\
         
       \thanks{T. Q. Duong is with the Faculty of Engineering and Applied Science, Memorial University, St. John’s, NL A1C 5S7, Canada and also with the School of Electronics, Electrical Engineering and Computer Science, Queen’s University Belfast, Belfast, U.K. (e-mail: tduong@mun.ca.)}\\

        }

	
	\maketitle

\begin{abstract}
This paper investigates continuous wideband orthogonal frequency-division multiplexing (OFDM) reception using an auxiliary (AUX)-assisted five-level Rydberg atomic quantum receiver (RAQR) with a single microwave local oscillator (LO), without requiring additional atomic reception transitions or LO frequency combs. By exploiting Floquet theory, the proposed framework characterizes the dynamic atomic response at finite signal-LO detunings without relying on a quasi-static Taylor expansion around the LO operating point. Specifically, closed-form radio-frequency (RF)-to-electrical conversion expressions are derived from the Lindblad master equation under weak-probe and weak-signal approximations. Building on these expressions, an equivalent multicarrier communication model is established, revealing the inherent frequency selectivity of the atomic response and the resulting subcarrier-dependent amplitude and phase distortions. RAQR-aware zero-forcing (ZF) equalization is developed to enable reliable OFDM reception. Furthermore, analytical results reveal how LO and AUX driving influence the atomic response through dressed-state splitting and energy-level shifts. \textcolor{black}{An upper bound on the LO-referenced half-power span is derived, and an approximately $18$-dB reduction in center-response power per bandwidth doubling is established in the AUX-free strong-LO regime. Simulation results validate the analytical response model and reveal pronounced subcarrier-dependent amplitude and phase variations across over $10$~MHz}, highlighting the potential of RAQRs as wideband receiving front ends for future wireless communications.
\end{abstract}


\section{Introduction}

Sixth-generation (6G) mobile networks are expected to support immersive
services, ubiquitous connectivity, and integrated sensing and
communication~\cite{wang2023road,peng2026rfparadigmrydbergatomic,10496996}. Realizing these
goals requires wider signal bandwidths and reliable weak-signal
reception within practical hardware and power
budgets~\cite{Tataria2021Wireless6G}. In conventional RF receivers, improving sensitivity may require a
larger hardware implementation and higher power consumption, while wider bandwidth increases integrated noise power, making sensitivity
harder to maintain within fixed hardware and power
budgets~\cite{Lin2015WidebandReceivers,Sharma2020ReceiverTradeoffs}. These challenges motivate
the exploration of alternative physical mechanisms for wireless
reception that complement continued advances in RF circuit design.

Rydberg atomic quantum receivers (RAQRs) offer an alternative reception
mechanism by converting RF-induced changes in atomic coherence into
optical signals~\cite{Gong2025WirelessOverview}. The large transition
dipole moments between Rydberg states enable sensitive interactions
with incident RF fields~\cite{Sedlacek2012Electrometry}. Specifically, a resonant probe beam is strongly absorbed without a
coupling laser, whereas adding the coupling laser under two-photon
resonance suppresses absorption and establishes a narrow transmission window, which is known as electromagnetically induced transparency
(EIT)~\cite{Mohapatra2007RydbergEIT}. Introducing a sufficiently strong
RF field resonant with the transition between two Rydberg states then
splits this EIT window into two resolved peaks, producing Autler–Townes
splitting (ATS) \cite{saglamyurek2018coherent}. Since the peak separation is proportional to the
RF-field amplitude, this spectral splitting provides an optical means
of measuring the field magnitude. To recover both amplitude and phase of the RF signal, atomic superheterodyne reception introduces a microwave local oscillator (LO) with a small frequency offset relative to the received signal, forming a time-varying optical readout. This superheterodyne architecture enables coherent RF reception for wireless communication networks~\cite{Jing2020Superheterodyne}.

Building on these detection mechanisms, RAQR research has progressed
from experimental demonstrations of waveform recovery to analytical
modeling of communication systems. Early experiments demonstrated
reception of amplitude- and frequency-modulated microwave
signals~\cite{Meyer2018Digital,Anderson2021AMFM}. Subsequent work investigated phase-shift-keyed
reception~\cite{Holloway2019PhaseMod}, quadrature amplitude modulation
(QAM) reception~\cite{Nowosielski2024QAM}, and phase-modulated
frequency-division multiplexing~\cite{Cai2023PhaseFDM}. To improve robustness against receiver impairments, laser phase-noise
compensation was incorporated into optically biased coherent
detection~\cite{Borowka2025OpticalBias}. In parallel, frequency-hopping
reception was investigated to enhance resilience to jamming by
exploiting atomic frequency tunability~\cite{Nan2026FHSS}.
Although these demonstrations establish the feasibility of atomic
reception, translating this capability into system-level design
requires tractable models that link incident electromagnetic signals
to the resulting optical responses. To this end, an equivalent baseband model was derived using a Taylor
expansion~\cite{Gongtcom2026}. Subsequently the linear operating regime of this approximate
model was characterized in~\cite{ChenTWC2026}. Building on this model, RAQR-based reception has been investigated in
self-sustained networks~\cite{pengJSAC2026} and satellite
communications~\cite{PengWCM2026}. These studies highlighted the potential of RAQRs as an alternative receiver technology for next-generation wireless networks.

Despite these advantages, high-sensitivity RAQR reception requires
highly stable laser driving and a tightly controlled atomic environment.
Fluctuations in laser detuning and power, together with atomic thermal
motion, would significantly alter the atomic susceptibility and RF-to-optical conversion gain~\cite{Schlossberger2026Linewidth}. Furthermore, for wideband communications, conventional resonance-based RAQRs face
two distinct constraints, namely limited instantaneous bandwidth and
discontinuous high-sensitivity carrier coverage. The former is governed
by optical driving and atomic relaxation
dynamics~\cite{Bohaichuk2022Transient,Hu2023Bandwidth}, whereas the latter
arises from the discrete RF transitions between Rydberg
states~\cite{Berweger2023Tuning}.  Therefore, broad carrier-frequency tunability
does not guarantee simultaneous reception over a continuous band.
These limitations motivate efforts to extend high-sensitivity carrier
coverage and broaden the instantaneous response bandwidth for
wideband multicarrier reception.

Several approaches have extended the frequency coverage and instantaneous bandwidth of RAQRs. Multilevel schemes enabled simultaneous reception through distinct Rydberg transitions, but their responses remained concentrated around separated resonances~\cite{Meyer2023Multiband,Cui2026Multiband}. To achieve continuous multicarrier reception, a five-level scheme combining auxiliary-field detuning with a microwave LO frequency comb was proposed in~\cite{Wang2026Multicarrier}, although comb generation and careful intermediate-frequency allocation increased implementation complexity. Alternatively, six-wave mixing was investigated to broaden the RF-to-optical response through multiphoton interactions~\cite{Yang2024Instantaneous,ChenJSSAC}. More recently, auxiliary-microwave control was introduced to enhance the instantaneous response of a five-level configuration~\cite{yan2026enhancinginstantaneousbandwidthrydberg}. Despite these advances, understanding how the broadened atomic response supports continuous multicarrier reception with a single LO, while accounting for its inherent frequency selectivity and gain-bandwidth tradeoff, remains an important challenge.

\textcolor{black}{Although the AUX-assisted five-level configuration in~\cite{yan2026enhancinginstantaneousbandwidthrydberg} demonstrated wideband microwave sensing, its broadband atomic response does not directly guarantee reliable multicarrier reception. In particular, real-valued IF readout introduces image overlap, while the frequency-selective atomic response causes subcarrier-dependent amplitude and phase distortions. To address these communication-specific challenges, we develop a single-LO OFDM reception framework without requiring an LO frequency comb. By integrating the Floquet-based atomic response with coherent optical detection, an equivalent multicarrier communication model and RAQR-aware equalizers are established. Beyond signal recovery, analytical results characterize the effects of LO and AUX driving on the center conversion gain and usable bandwidth, revealing a fundamental gain--bandwidth tradeoff. The main contributions are summarized as follows:}
\begin{itemize}

\item \textbf{Floquet-based wideband conversion model.}
Building on the first-order Floquet response of the AUX-assisted five-level configuration, we derive a communication-oriented RF-to-electrical transfer function incorporating coherent optical readout. Unlike quasi-static Taylor expansions around the LO operating point, the proposed model captures finite-frequency atomic dynamics and the resulting amplitude and phase responses. Simulation results validate the derived expressions and demonstrate a one-sided \textcolor{black}{LO-referenced} bandwidth over $11~\mathrm{MHz}$ under the considered parameters, highlighting the potential of RAQRs as wideband receiving front ends.

\item \textbf{Single-LO wideband OFDM reception.}
We extend the proposed reception scheme to wideband OFDM using a single microwave LO, without an LO frequency comb. An equivalent fast Fourier transform (FFT)-domain input-output model is derived to characterize the inherently frequency-selective atomic response, which introduces subcarrier-dependent amplitude and phase distortions even under a frequency-flat wireless channel. To address these distortions, we develop RAQR-aware frequency-domain equalizers that jointly compensate for the wireless and atomic responses. 

\item \textbf{Analytical characterization of LO/AUX response control.}
Under the far-detuned AUX approximation, a reduced complex-response model is derived to characterize the resonance locations, modal linewidths, and
center atomic conversion gain. The analysis reveals that LO
driving controls the dressed-state splitting and linewidths,
whereas AUX driving shifts the bright-state resonances
and modifies the center response through distinct energy-level
shifts. We further establish an explicit condition for AUX-induced gain enhancement, providing analytical guidance
for shaping the atomic frequency response.

\item \textbf{Bandwidth upper bound and gain--bandwidth tradeoff.}
We derive an analytical upper bound on the one-sided \textcolor{black}{LO-referenced} bandwidth and establish a quantitative gain-bandwidth tradeoff for the AUX-free case. In the strong-LO regime with a contiguous passband, doubling the bandwidth incurs an approximately $18$-dB loss in center-response power, implying an equivalent sensitivity penalty under fixed readout-noise conditions. Simulation results validate the bandwidth bound and corroborate the analyzed gain-bandwidth tradeoff.

\end{itemize}

The remainder of this paper is organized as follows.
Section~II presents the five-level atomic model, and Section~III derives the optical response and its equivalent multicarrier representation.
Section~IV analyzes the tradeoff between gain and bandwidth.
Section~V presents simulation results, and Section~VI concludes the paper.

\section{System Model}

\subsection{Five-Level Floquet–Lindblad Model for Atomic Response}

\begin{figure}
    \centering
    \includegraphics[width=0.9\linewidth]{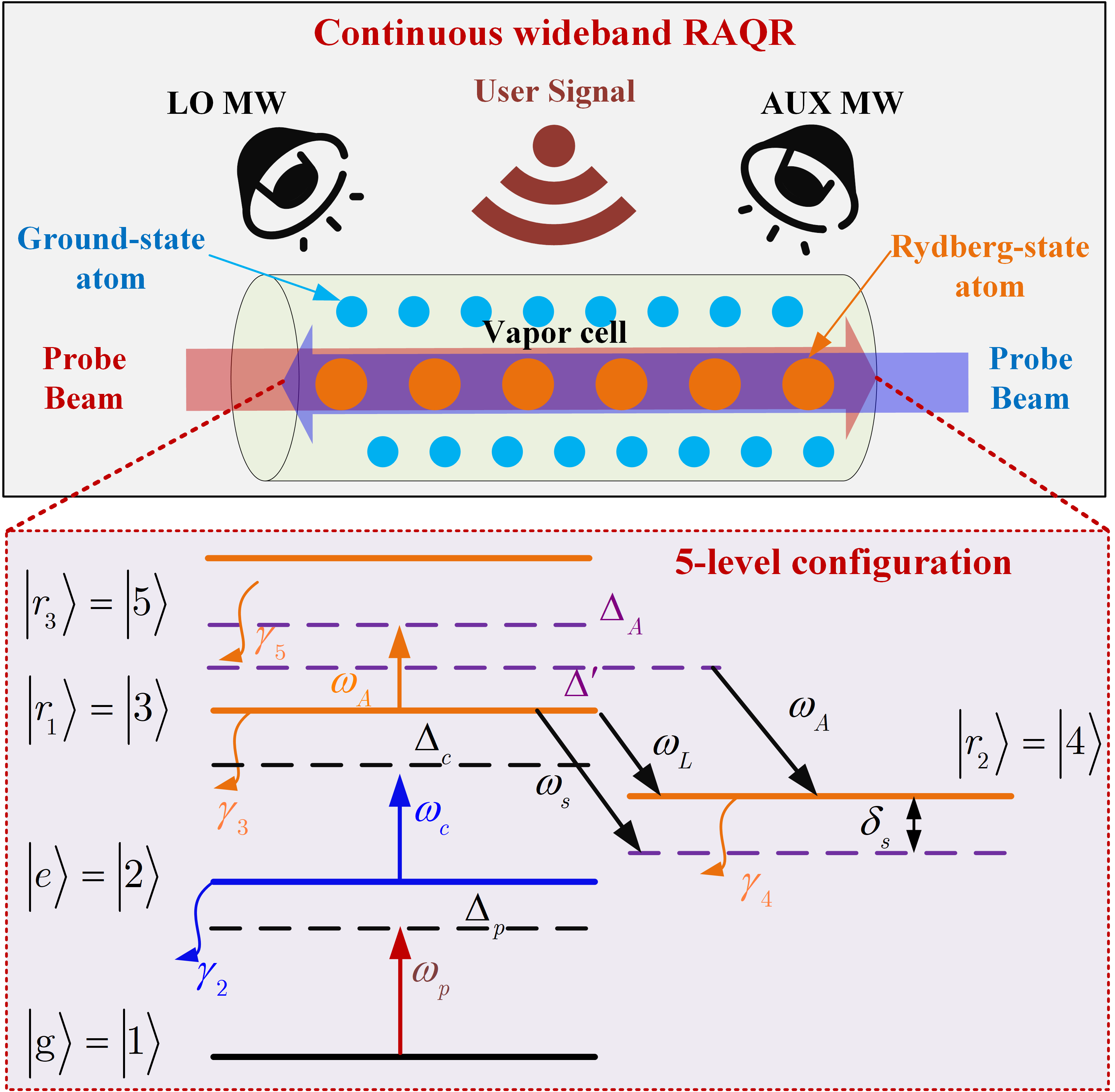}
    \caption{Continuous wideband RAQR based on a 5-level configuration.}
    \label{5level}
\end{figure}
In Fig.~\ref{5level}, we consider a five-level configuration with the basis \(\left\{
|1\rangle,|2\rangle,|3\rangle,|4\rangle,|5\rangle
\right\}\),
where \(|1\rangle\) is the ground state, \(|2\rangle\) represents the excited state, and \(|3\rangle\), \(|4\rangle\), and \(|5\rangle\) are Rydberg states. The probe and coupling lasers drive $|1\rangle\leftrightarrow|2\rangle$ and $|2\rangle\leftrightarrow|3\rangle$ with Rabi frequencies $\Omega_p$ and $\Omega_c$. The LO and weak RF signal drive $|3\rangle\leftrightarrow|4\rangle$ with Rabi amplitudes $\Omega_L$ and $\Omega_s$, while the AUX field drives $|3\rangle\leftrightarrow|5\rangle$ with $\Omega_A$. The signal-LO angular-frequency offset is \(\delta_s=\omega_s-\omega_L\), where \(\omega_s\) and \(\omega_L\) are the angular frequencies of the signal MW and the LO MW, respectively. The angular frequencies of probe, coupling, and AUX MW are denoted as \(\omega_p\), \(\omega_c\), and \(\omega_A\), respectively. 
Furthermore, the constant phases of the probe, coupling, LO, and AUX
fields are absorbed into the atomic basis, allowing their Rabi
frequencies to be taken as real and nonnegative. 

By applying the rotating-wave approximation, the Hamiltonian \(\mathbf H(t)\) is written as
\begin{equation}
\begin{aligned}
&{\mathbf H(t)}=\hbar\times\\
&\begin{bmatrix}
0 & \frac{\Omega_p}{2} & 0 & 0 & 0\\
\frac{\Omega_p}{2} & -\Delta_p & \frac{\Omega_c}{2} & 0 & 0\\
0 & \frac{\Omega_c}{2} & -D_3 &
\frac{\Omega_L+\Omega_{s}e^{-{\rm j}(\delta_s t+\theta_s)}}{2}
& \frac{\Omega_A}{2}\\
0 & 0 &
\frac{\Omega_L+\Omega_{s}e^{{\rm j}(\delta_s t+\theta_s)}}{2}
& -D_4 & 0\\
0 & 0 & \frac{\Omega_A}{2} & 0 & -D_5
\end{bmatrix},
\end{aligned}
\label{eq:H_full}
\end{equation}
where \(\hbar\) denotes the reduced Planck constant.
Here, \(\theta_s = \phi_s - \phi_{\rm LO}\) is the phase shift between the MW signal (\(\phi_s\)) and LO (\(\phi_{\rm LO}\)).   \(D_3\), \(D_4\), and \(D_5\) are given by
\begin{equation}
\label{D_3}
D_3
=
\Delta_p+\Delta_c+\xi_A,
\end{equation}
\begin{equation}
\label{D_4}
D_4
=
\Delta_p+\Delta_c-\Delta_L-\xi_A,
\end{equation}
and
\begin{equation}
\label{D_5}
D_5
=
\Delta_p+\Delta_c+\Delta_A,
\end{equation} 
where $\xi_A=\frac{\mu_{34}^2}{\mu_{35}^2}\frac{\Omega_A^2}{4\Delta'}$ is the AUX-MW-induced alternating current (AC) Stark shift\footnote{Since the AUX field couples $|3\rangle\leftrightarrow|4\rangle$ off-resonantly with the Rabi frequency $\frac{\mu_{34}}{\mu_{35}}\Omega_A$, this shift carries the factor $\mu_{34}^2/\mu_{35}^2$}, and \(\mu_{i,j}\) denotes the dipole moment between level \(|i\rangle\) and \(|j\rangle\). \(\Delta_p = \omega_p-\omega_{21}\), \(\Delta_c=\omega_c-\omega_{32}\), \(\Delta_L=\omega_L-\omega_{34}\), and \(\Delta_A=\omega_A-\omega_{53}\) denote the detunings of the probe laser, coupling laser, LO MW, and AUX MW, respectively. \(\Delta'=\omega_A-\omega_{34}
=\Delta_A+\omega_{53}-\omega_{34}\) denotes the detuning of the AUX MW with respect to the \(|3\rangle\leftrightarrow |4\rangle\) transition.

From \eqref{eq:H_full}, the Hamiltonian can be decomposed as
\begin{equation}
\mathbf H(t)
=
\mathbf H_0
+
\Omega_{s}\mathbf H_+e^{{\rm j}(\delta_s t+\theta_s)}
+
{\Omega_s}\mathbf H_-e^{-{\rm j}(\delta_s t+\theta_s)},
\label{eq:H_decompose_phase}
\end{equation}
where \(\mathbf{H}_0\), \(\mathbf H_+\) and \(\mathbf H_-\) are expressed as
\begin{equation}
\mathbf{H}_0
=
\hbar\begin{bmatrix}
0 & \frac{\Omega_p}{2} & 0 & 0 & 0\\
\frac{\Omega_p}{2} & -\Delta_p & \frac{\Omega_c}{2} & 0 & 0\\
0 & \frac{\Omega_c}{2} & -D_3 & \frac{\Omega_L}{2} & \frac{\Omega_A}{2}\\
0 & 0 & \frac{\Omega_L}{2} & -D_4 & 0\\
0 & 0 & {\frac{\Omega_A}{2}} & 0 & -D_5
\end{bmatrix},
\label{eq:H0_main}
\end{equation}
\begin{equation}
\mathbf{H}_{+}
=
\frac{\hbar}{2}|4\rangle\langle 3|,
\end{equation}
and
\begin{equation}
\mathbf{H}_{-}
=
\frac{\hbar}{2}|3\rangle\langle 4|.
\end{equation}
Therefore, the LO MW is included in the time-independent part \(\mathbf{H}_0\), while the received signal MW appears as a time-periodic perturbation at the intermediate angular frequency \(\delta_s\).

Let \(\boldsymbol{\rho}(t)\in\mathbb{C}^{5\times5}\) denote the density matrix of the atomic system and  \(\rho_{ij}\) mean the \((i,j)\)-th element of \(\boldsymbol{\rho}(t)\).
The density matrix is governed by the Lindblad master equation \cite{yan2026enhancinginstantaneousbandwidthrydberg}
\begin{equation}
\begin{split}
\frac{d\boldsymbol\rho}{dt}
&=-\frac{{\rm j}}{\hbar}[\mathbf H(t),\boldsymbol\rho]
-\frac12\{\boldsymbol\Gamma,\boldsymbol\rho\}
+\boldsymbol\Lambda,\\
\boldsymbol\Gamma
&=\operatorname{diag}\{\gamma,\gamma+\gamma_2,
\gamma+\gamma_r,\gamma+\gamma_r,\gamma+\gamma_r\},\\
\boldsymbol\Lambda
&=\operatorname{diag}\{\gamma+\gamma_2\rho_{22}
+\gamma_r\rho_{44},\gamma_r\rho_{33},\gamma_r\rho_{55},0,0\}.
\end{split}
\label{mastereq}
\end{equation}
where \(\boldsymbol{\Gamma}\) is the relaxation matrix and
\(\boldsymbol{\Lambda}\) is the repopulation matrix. $\gamma_2$ denotes the spontaneous population decay rate of state $|2\rangle$, \(\gamma_r\) is a phenomenological approximation for the three Rydberg population decay rates, and $\gamma$ represents \textcolor{black}{the effective dephasing rate, including the transit broadening, the residual two-photon Doppler dephasing, and the dipole--dipole broadening}.

To obtain the response at the intermediate frequency \(\delta_s\), we use the Floquet theory. Under the weak-signal condition, i.e., \(|\Omega_s|\ll |\Omega_L|\),
the density matrix is approximated by\footnote{Higher-order sideband components of the density matrix are neglected,
as they are of second or higher order in $\Omega_s$.}
\begin{equation}
\boldsymbol{\rho}(t)
\approx
\boldsymbol{\rho}^{(0)}
+
\boldsymbol{\rho}^{(+1)}e^{\mathrm{j}(\delta_s t +\theta_s)}
+
\boldsymbol{\rho}^{(-1)}e^{-\mathrm{j}(\delta_s t +\theta_s)},
\label{eq:R_floquet_main}
\end{equation}
where \(\boldsymbol{\rho}^{(0)}\) is the zeroth-order density matrix induced by the probe laser, coupling laser, LO MW, and AUX MW, while \(\boldsymbol{\rho}^{(+1)}\) and \(\boldsymbol{\rho}^{(-1)}\) are the first-order sideband density matrices induced by the weak signal MW. Then, the power of the output probe beam is related to \(\rho_{21}\), which is given by
\begin{equation}\label{outprobe}
\begin{split}
        P_{\text{out}}(\rho_{21}) &= P_{\text{in}}  \exp\Bigg\{L_\text{cell} k_p D_{\Omega}\Im\Big[\rho_{21}(t)\Big] \Bigg\},
\end{split}
\end{equation}
where \(P_\text{in}\) means the power of the input probe beam and \(L_{\rm cell}\) is the length of vapor cell.  \(D_{\Omega} = \frac{2 N_0 \mu^2_{12}}{\epsilon_0\hbar \Omega_p}\) represents the constant term, $N_0$ is the atomic number density, $k_p=2\pi/\lambda_p$ is the probe wavenumber,
and $\mu_{12}$ is the effective dipole moment.

\section{Wideband Reception Model}
This section derives approximate closed-form expressions
for the optical response using Floquet theory and then
develops the corresponding wideband multicarrier
reception model.
\subsection{Output Probe Beam}
As can be seen from \eqref{mastereq}, it is challenging to derive the exact expression of the output probe beam. To tackle this issue,  we derive the closed-form expression of \(\rho_{21}\) under the following physically motivated assumptions. Particularly, we assume that the system operates in the weak-probe regime \cite{yan2026enhancinginstantaneousbandwidthrydberg}, i.e., \(\rho_{11} \approx 1\), \(\rho_{ii} \approx 0\), \(i = 2,3,4,5\). Based on these assumptions, the optical coherence can be expanded as
\begin{equation}\label{rho21t}
\rho_{21}(t)
\approx
\rho_{21}^{(0)}
+
\rho_{21}^{(+)}e^{{\rm j}(\delta_s t+\theta_s)}
+
\rho_{21}^{(-)}e^{-{\rm j}(\delta_s t+\theta_s)}.
\end{equation}
Then, the closed-form expression of the steady-state optical coherence is given by
\begin{equation}
\label{eq:rho21_0_closed_main}
\rho_{21}^{(0)}
=\frac{\Omega_p\Xi(0)}{2F(0)},\qquad
\rho_{21}^{(\pm)}=\Omega_sH_\pm(\delta_s),
\end{equation}
where
\begin{equation}
H_+(\delta_s)=\frac{\Omega_p\Omega_c^2\Omega_L}
{32F(0)Z_4(\delta_s)F(\delta_s)},
\label{eq:rho21_plus_closed_main}
\end{equation}
\begin{equation}
H_-(\delta_s)=\frac{\Omega_p\Omega_c^2\Omega_L}
{32F(0)Z_4(0)F(-\delta_s)}.
\label{eq:rho21_minus_closed_main}
\end{equation} 
Furthermore, \(\Xi(\delta_s)\) and \(F(\delta_s)\) are given by
\begin{equation}
    \Xi(\delta_s)=Z_3(\delta_s)-\frac{\Omega_L^2}{4Z_4(\delta_s)}
-\frac{\Omega_A^2}{4Z_5(\delta_s)},
\end{equation}
and 
\begin{equation}
    F(\delta_s)=Z_2(\delta_s)\Xi(\delta_s)-\frac{\Omega_c^2}{4},
\end{equation}
where 
\begin{equation}
\begin{split}
    Z_2(\delta_s)&=\Delta_p-\delta_s+{\rm j}\eta_2,\\
Z_k(\delta_s)&=D_k-\delta_s+{\rm j}\eta_r,\quad k=3,4,5.
\end{split}
\label{eq:XiF}
\end{equation}
In \eqref{eq:XiF}, \(\eta_2=\gamma+\frac{\gamma_2}{2}\) and \(\eta_r=\gamma+\frac{\gamma_r}{2}\). The detailed derivation is provided in
Appendix~\ref{app:weak_probe_closed_form}.

By using the first-order coherence approximation and substituting \eqref{rho21t} into \eqref{outprobe}, we have
\begin{equation}
\begin{split}
    P_{\rm out}(t)
&\approx
P_{\rm out}^{(0)}
\exp\Big\{
L_{\rm cell}k_pD_{\Omega}\\
&\times \Im\big\{
\rho_{21}^{(+)}e^{{\rm j}(\delta_s t+\theta_s)}
+
\rho_{21}^{(-)}e^{-{\rm j}(\delta_s t+\theta_s)}
\big\}
\Big\}, 
\end{split}
\end{equation}
where \(P_{\rm out}^{(0)}\) is given by
\begin{equation}
\label{eq:Pout0_power}
P_{\rm out}^{(0)}
=
P_{\rm in}
\exp\left\{
L_{\rm cell}k_pD_{\Omega}
\Im\big\{
\rho_{21}^{(0)}
\big\}
\right\}.
\end{equation}
Since \(\rho_{21}^{(+)}\) and \(\rho_{21}^{(-)}\) are first-order small
quantities induced by the weak signal field, the exponential term can be
linearized by using \(e^{x}=1+x+\mathcal O(x^2)\). Therefore, we have
\begin{equation}\label{pout0}
\begin{split}
    P_{\rm out}(t)
&\approx
P_{\rm out}^{(0)}
\Bigg\{
1
+
L_{\rm cell}k_pD_{\Omega}\Omega_s
\Im\Big\{
H_+(\delta_s)\\
&\times e^{{\rm j}(\delta_s t+\theta_s)}
+
H_-(\delta_s)e^{-{\rm j}(\delta_s t+\theta_s)}
\Big\}
\Bigg\}.
\end{split}
\end{equation}
Meanwhile, the corresponding phase of the transmitted probe beam is
approximated as
\begin{equation}
\label{eq:probe_phase_bcod}
\begin{split}
\phi_{\rm p}(t)
&\approx
\phi_{\rm p}^{(0)}
-
\frac{L_{\rm cell}k_pD_{\Omega}\Omega_s}{2}
\\
&\times
\Re\Big\{
H_+(\delta_s)e^{{\rm j}(\delta_st+\theta_s)}
+
H_-(\delta_s)e^{-{\rm j}(\delta_st+\theta_s)}
\Big\},
\end{split}
\end{equation}
where \(\phi_{\rm p}^{(0)}
=
\phi_{\rm p,in}
-
\frac{L_{\rm cell}k_pD_{\Omega}}{2}
\Re\left\{\rho_{21}^{(0)}\right\}\) and \(\phi_{\rm p,in}\) is the phase of the input probe beam.

\subsection{Equivalent Model}

To improve the optical readout sensitivity, balanced coherent optical
detection (BCOD) is adopted \cite{Gongtcom2026}. The transmitted
probe beam can be expressed as
\begin{equation}
x_{\rm p}(t)
=
\sqrt{2P_{\rm out}(t)}
\cos\left[
\omega_pt+\phi_{\rm p}(t)
\right].
\end{equation}
A phase-coherent reference optical beam is introduced as
\begin{equation}
x_{\rm r}(t)
=
\sqrt{2P_{\rm r}}
\cos\left(
\omega_pt+\phi_{\rm r}
\right),
\end{equation}
where \(P_{\rm r}\) and \(\phi_{\rm r}\) denote the power and phase of
the reference optical beam, respectively. 

After combining the transmitted probe and the reference optical beam
through a beam splitter and subtracting the photocurrents of the
two photodetectors, the BCOD output voltage is given by
\begin{equation}
\label{eq:bcod_output_general}
V_{\rm B}(t)
=
2\sqrt{G_{\rm A}}R_{\rm L}\mathcal R_{\rm pd}
\sqrt{
P_{\rm r}P_{\rm out}(t)
}
\cos
\left[
\phi_{\rm r}-\phi_{\rm p}(t)
\right],
\end{equation}
{Following the electrical-gain convention in~\cite{Gongtcom2026}, \(G_{\rm A}\) is the linear post-detection power gain. The quantities \(\mathcal R_{\rm pd}\) and \(R_{\rm L}\) are the photodetector responsivity and load resistance, respectively.} For simplicity, we assume that the reference beam is phase aligned with the steady-state
transmitted probe, i.e., \(\phi_{\rm r}=\phi_{\rm p}^{(0)}\). Consequently, 
\(\cos[\phi_{\rm r}-\phi_{\rm p}(t)]
=1+\mathcal O(\Omega_s^2)\) can be approximated as 1, under the weak-signal condition, i.e., \(\Omega_s\ll\Omega_L\). Then, substituting
\eqref{pout0} and \eqref{eq:probe_phase_bcod} into
\eqref{eq:bcod_output_general} and retaining only the first-order terms
with respect to \(\Omega_s\), we have
\begin{equation}
V_{\rm B}(t)
\approx
V_{\rm DC}
+
V_{\rm AC}(t),
\end{equation}
where
\begin{equation}
V_{\rm DC}
=
2\sqrt{G_{\rm A}}R_{\rm L}\mathcal R_{\rm pd}
\sqrt{
P_{\rm r}P_{\rm out}^{(0)}
},
\end{equation}
and
\begin{equation}
\label{eq:vac_bcod}
\begin{split}
V_{\rm AC}(t)
=&
\sqrt{G_{\rm A}}R_{\rm L}\mathcal R_{\rm pd}
\sqrt{
P_{\rm r}P_{\rm out}^{(0)}
}
L_{\rm cell}k_pD_{\Omega}\Omega_s
\\
&\times
\Im\Big\{
\left[
H_+(\delta_s)-H_-^*(\delta_s)
\right]
e^{{\rm j}(\delta_st+\theta_s)}
\Big\}.
\end{split}
\end{equation}

We next derive an idealized steady-state communication model from the output voltage. The symbol interval of duration is
\(T_s\), during which the wireless channel coefficient and transmitted
symbol are assumed to remain constant. Then, the Rabi frequency \(\Omega_s
= \frac{\mu_{34}}{\hbar}E_s\) is determined by the amplitude of the
received signal MW \(E_s =
\frac{\sqrt{60P_tG_t}}{d_{\rm dis}}
|hs|\), where \(P_t\), \(G_t\), and \(d_{\rm dis}\) denote the transmit
power, transmit antenna gain, and propagation distance, respectively. \(h\sim  \mathcal{CN}(0,1)\) and \(s \in \mathbb{C}^{1\times 1}\) denote the wireless channel coefficient and transmitted information during the current symbol interval. Furthermore, the received signal phase and its phase relative to the LO MW are \(\phi_s
=
\arg\{hs\}\) and \(\theta_s = \phi_s-\phi_{\rm LO}\)
respectively. Substituting these results into \eqref{eq:vac_bcod}, we have
\begin{equation}
    \begin{split}
          V_{\rm AC}(t)
& =\sqrt{G_{\rm A}}R_{\rm L}\mathcal R_{\rm pd}
\sqrt{
P_{\rm r}P_{\rm out}^{(0)}
}
L_{\rm cell}k_pD_{\Omega}\frac{\mu_{34}}{\hbar}\frac{\sqrt{60P_t G_t}}{d_{\rm dis}} \\
&\times|hs| \Im\Big\{\big[H_+(\delta_s) -H^*_-(\delta_s) \big] e^{j(\delta_st+\theta_s)} \Big\}.
    \end{split}
\end{equation}

Then, the imaginary part of \(\Big[H_+(\delta_s) -H^*_-(\delta_s) \Big] e^{j(\delta_st+\theta_s)}\) can be rewritten as
\begin{equation}
\begin{split}
        &\Im\Big\{\big[H_+(\delta_s) -H^*_-(\delta_s) \big] e^{j(\delta_st+\theta_s)} \Big\}\\
         =& \Re\{H_+(\delta_s) -H^*_-(\delta_s) \}\sin(\delta_s t+\theta_s) \\
        +&  \Im\{H_+(\delta_s) -H^*_-(\delta_s) \}\cos(\delta_s t+\theta_s).
\end{split}
\end{equation}
We assume ideal coherent downconversion followed by low-pass filtering to reject the image at angular frequency $2|\delta_s|$. For the finite-interval projection below, a nonzero IF satisfying $\delta_sT_s=2\pi\ell$, $\ell\in\mathbb Z\setminus\{0\}$, is selected, and thus the image term integrates to zero over the symbol interval. The recovered complex baseband symbol is given by
\begin{equation}
\begin{split}
        &\frac{2}{T_s}\int_0^{T_s}V_{\rm AC}(t)\sin(\delta_s t)dt
+
{\rm j}
\frac{2}{T_s}\int_0^{T_s}V_{\rm AC}(t)\cos(\delta_s t)dt\\
        = & g_{\rm c}\sqrt{P_t}\left[
H_+(\delta_s)-H_-^*(\delta_s)
\right]hse^{-\mathrm j \phi_{\rm LO}},
\end{split}
\end{equation}
where \(g_{\rm c} =\sqrt{G_{\rm A}}R_{\rm L}\mathcal R_{\rm pd}
\sqrt{
P_{\rm r}P_{\rm out}^{(0)}
}
L_{\rm cell}k_pD_{\Omega}\frac{\sqrt{60G_t}}{d_{\rm dis}}\frac{\mu_{34}}{\hbar}\) denotes the conversion gain. Therefore, the received symbol is 
\begin{equation}
\begin{split}
      y &= g_{\rm c}\sqrt{P_t}\left[
H_+(\delta_s)-H_-^*(\delta_s)
\right]e^{-\mathrm j \phi_{\rm LO}}hs + w \\
& \triangleq \sqrt{P_t}g(\delta_s)hs + w
\end{split}
\end{equation}
where \(g(\delta_s) =g_{\rm c}\left[
H_+(\delta_s)-H_-^*(\delta_s)
\right]e^{-\mathrm j \phi_{\rm LO}} \). Following~\cite{Gongtcom2026}, we model the aggregate quantum projection
noise, photon shot noise, and intrinsic thermal noise at the demodulator
output as equivalent complex Gaussian noise,
$w\sim\mathcal{CN}(0,\sigma^2)$.

\subsection{Multicarrier Extension}
\label{subsec:multi_carrier_extension}

We next extend the single-carrier model to an idealized one-sided OFDM
signal with $N_c$ subcarriers and a useful symbol duration of $T_s$ \cite{zhu2009chunk,zhu2011chunk}.
Consistent with the preceding ideal-detection assumptions, cyclic-prefix
overhead, atomic memory across symbol boundaries, and transient effects
are neglected. Let
\(n\in\mathcal N_c\triangleq\{1,\cdots,N_c\}\) and \(\delta_s^n = \omega_s^n-\omega_L\) denote the subcarrier 
index and the angular-frequency offset between the \(n\)-th signal
subcarrier and the LO MW, respectively. Similar to the single-carrier case, we consider an arbitrary multicarrier symbol interval of duration \(T_s\) and assume that the wireless channel remains approximately constant within this interval. Consequently, the channel coefficient \(h_n\) and transmitted symbol \(s_n\) are treated as constants over the considered symbol interval, and the symbol index is omitted for notational simplicity.

The phase shift between the received signal of the \(n\)-th subcarrier and the LO MW is \(  \theta_n = \phi_{s,n}-\phi_{\rm LO}\), where \(\phi_{s,n}= \arg\{h_ns_n\}\). Then, the Rabi-frequency amplitude is written as
\begin{equation}
\label{eq:Omega_multicarrier}
\Omega_{s,n}
=
\frac{\mu_{34}}{\hbar}
\frac{\sqrt{60P_{t,n}G_t}}{d_{\rm dis}}
\left|h_n s_n\right|,
\end{equation}
{where \(P_{t,n}\) denotes the average power allocated to subcarrier \(n\), with \(\mathbb E|s_n|^2=1\) and \(\sum_{n=1}^{N_c}P_{t,n}=P_t\). Here,} \(h_n\) and \(s_n\) denote the channel coefficient and transmitted symbol on the \(n\)-th subcarrier, respectively. Accordingly, the signal contribution to the rotating-frame Hamiltonian is obtained by superposing the contributions from all subcarriers.

Based on the above discussion, the multicarrier Hamiltonian can be
written as
\begin{equation}
\label{eq:H_multicarrier_decomposition}
\begin{split}
\mathbf H(t)
&=
\mathbf H_0
+
\sum_{n=1}^{N_c}
\Omega_{s,n}
\Big[
\mathbf H_{+}
e^{{\rm j}(\delta_s^n t+\theta_n)}\\
&+
\mathbf H_{-}
e^{-{\rm j}(\delta_s^n t+\theta_n)}
\Big].
\end{split}
\end{equation}
We assume that the aggregate multicarrier microwave field is weak
relative to the LO field. Under the first-order approximation, each subcarrier independently induces sideband coherences at \(\pm\delta_s^n\), while the zeroth-order coherence \(\rho_{21}^{(0)}\) remains unchanged and higher-order intermodulation terms are neglected. Solving the Lindblad master
equation under these approximations, we obtain
\begin{equation}
\label{eq:rho21_multicarrier}
\begin{split}
\rho_{21}(t)
&\approx
\rho_{21}^{(0)} +
\sum_{n=1}^{N_c}
\Omega_{s,n}
H_{+}(\delta_s^n)
e^{{\rm j}(\delta_s^n t+\theta_n)}
\\
&+
\sum_{n=1}^{N_c}
\Omega_{s,n}
H_{-}(\delta_s^n)
e^{-{\rm j}(\delta_s^n t+\theta_n)}.
\end{split}
\end{equation}
Therefore, compared with the single-carrier case, the first-order
atomic response is simply the superposition of the sideband
responses generated by the \(N_c\) signal components.

Substituting the first-order coherence in \eqref{eq:rho21_multicarrier} into \eqref{outprobe}, we have
\begin{equation}
\label{eq:Pout_multicarrier_exact}
\begin{split}
P_{\rm out}(t)
&\approx
P_{\rm out}^{(0)}
\exp
\Bigg\{
L_{\rm cell}k_pD_{\Omega}
\sum_{n=1}^{N_c}
\Omega_{s,n}
\\
&\times
\Im\Big\{\Big[H_+(\delta^n_s) -H^*_-(\delta^n_s) \Big] e^{j(\delta^n_st+\theta_n)} \Big\}\Bigg\}.
\end{split}
\end{equation}
Under the same first-order weak-signal approximation, the exponential
term in \eqref{eq:Pout_multicarrier_exact} can be approximated as
\begin{equation}
\label{eq:Pout_multicarrier_linear}
\begin{split}
P_{\rm out}(t)
&\approx
P_{\rm out}^{(0)}
\Big\{
1
+
L_{\rm cell}k_pD_{\Omega}
\sum_{n=1}^{N_c}
\Omega_{s,n}
\\
&\times
\Im\big\{\Big[H_+(\delta^n_s) -H^*_-(\delta^n_s) \Big] e^{j(\delta^n_st+\theta_n)} 
\big\}\Big\}.
\end{split}
\end{equation}

After photodetection and removal of the DC component, the corresponding
AC voltage is
\begin{equation}
\label{eq:VAC_multicarrier}
\begin{split}
V_{{\rm AC},N_c}(t)
&=
\sqrt{G_{\rm A}}R_{\rm L}\mathcal R_{\rm pd}
\sqrt{
P_{\rm r}P_{\rm out}^{(0)}
}
L_{\rm cell}k_pD_{\Omega}\sum_{n=1}^{N_c}
\Omega_{s,n}
\\
&\times
\Im
\Big\{
\left[
H_{+}(\delta_s^n)
-
H_{-}^{*}(\delta_s^n)
\right]
e^{{\rm j}(\delta_s^n t+\theta_n)}
\Big\}.
\end{split}
\end{equation}

For notational convenience, we define the frequency-dependent complex
response of the RAQR as
\begin{equation}
\label{eq:g_multicarrier}
g(\delta_s^n)
=
g_{\rm c}
\left[
H_{+}(\delta_s^n)
-
H_{-}^{*}(\delta_s^n)
\right]
e^{-{\rm j}\phi_{\rm LO}},
\end{equation}
Then, the multicarrier AC voltage can be rewritten as
\begin{equation}
\label{eq:VAC_multicarrier_compact}
V_{{\rm AC},N_c}(t)
=
\sum_{n=1}^{N_c}
\Im
\left\{
\sqrt{P_{t,n}}\,
g(\delta_s^n)
h_ns_n
e^{{\rm j}\delta_s^n t}
\right\}.
\end{equation}
Equation~\eqref{eq:VAC_multicarrier_compact} shows that the
photodetector output is a real-valued multi-tone IF waveform, whose
\(n\)-th component has the complex envelope
\(\sqrt{P_{t,n}}g(\delta_s^n)h_ns_n\). Compared with the
single-carrier case, the RAQR therefore simultaneously maps the
received multicarrier RF signal onto multiple IF components with
subcarrier-dependent atomic amplitude and phase responses.

In the following, we describe the FFT-based digital demodulation of \(V_{{\rm AC},N_c}(t)\). First, we place the data subcarriers below the LO, namely $\omega_s^n=\omega_L-\frac{2\pi n}{T_s}$, and thus $\delta_s^n=\omega_s^n-\omega_L=-\frac{2\pi n}{T_s}$, $n\in\mathcal N_c$. Each symbol interval is uniformly sampled at $M>2N_c$ points,
corresponding to a sampling rate of $F_s=M/T_s$. The $m$-th sample
is given by
\begin{equation}
\label{eq:ADC_multicarrier}
\begin{split}
    v_{{\rm AC}}[m]
&=
V_{{\rm AC},N_c}
\big(\frac{mT_s}{M}
\big)
\\
&=\sum_{n=1}^{N_c}\Im\Big\{\sqrt{P_{t,n}}\,g(\delta_s^n)h_ns_n
e^{\mathrm j\frac{\delta_s^n mT_s}{M}}\Big\}
,
\end{split}
\end{equation}
Exploiting discrete Fourier transform (DFT) orthogonality, the noisy observation on the $n$th subcarrier is obtained through an $M$-point FFT as
\begin{equation}
\label{eq:FFT_demodulation}
\begin{split}
 y_n&=\frac{2\mathrm j}{M}\sum_{m=0}^{M-1}v_{{\rm AC}}[m]e^{-\mathrm j\frac{\delta_s^n m T_s}{M}}\\
&=\sqrt{P_{t,n}}\,g(\delta_s^n)h_ns_n+\hat w_n,
\end{split}
\end{equation}
where \(\hat w_n\sim \mathcal{CN}(0,\sigma^2_n)\) 
denotes the equivalent noise after FFT demodulation. \textcolor{black}{Following~\cite{Gongtcom2026}, the receiver-noise model incorporates quantum projection noise, photon shot noise, and thermal noise.}

Stacking the FFT outputs across all $N_c$ subcarriers
yields the vector model
\begin{equation}
\label{eq:FFT_RAQR_vector_model}
\mathbf y
=
\mathbf G(\boldsymbol{\delta}_s)
\mathbf H_c
\mathbf P_t^{1/2}
\mathbf s
+
\mathbf{\hat w},
\end{equation}
where \(\mathbf y
=
\left[
y_1,
\cdots,
y_{N_c}
\right]^{\rm T}\), \(\mathbf s
=
\left[
s_1,
\cdots,
s_{N_c}
\right]^{\rm T}\),  and \(\mathbf{\hat w}
=
\left[
\hat w_1,
\cdots,
\hat w_{N_c}
\right]^{\rm T}\) denote the received-signal, transmitted-symbol, and noise vectors, respectively. The transmit-power allocation matrix is defined as \(\mathbf P_t
=
\operatorname{diag}
\left\{
P_{t,1},
\cdots,
P_{t,N_c}
\right\}\). 
Furthermore, \(\mathbf G(\boldsymbol{\delta}_s)
=
\operatorname{diag}
\left\{
g(\delta_s^1),
\cdots,
g(\delta_s^{N_c})
\right\}\) and \(\mathbf H_c
=
\operatorname{diag}
\left\{
h_1,
\cdots,
h_{N_c}
\right\}\) collect the complex frequency responses and wireless channel coefficients across the \(N_c\) subcarriers, respectively.

\begin{remark}[One-sided bandwidth]
Although the RAQR exhibits responses at both positive and negative microwave-LO frequency offsets, the real-valued BCOD output maps the two sidebands onto the same demodulated frequency band. Therefore, the transmitted signals on both sides would overlap and create image interference. To avoid this ambiguity, data subcarriers are allocated to only one side of the LO. For an approximately symmetric response, the resulting usable bandwidth for independent data transmission is about half of the double-sided atomic-response bandwidth.
\end{remark}

\begin{remark}[Frequency-dependent atomic response]\label{remarkFDR}
Equation~\eqref{eq:FFT_RAQR_vector_model} shows that the RAQR first converts the received RF multicarrier signal into a real-valued multi-tone IF waveform, which is then digitized and decomposed into orthogonal subcarriers via a single FFT. However, the FFT only separates the orthogonal subcarriers and does not compensate for either the frequency-dependent RAQR response or the wireless propagation channel. In particular, the strongly frequency-dependent atomic response may cause substantial subcarrier-dependent amplitude and phase variations, motivating the RAQR-aware frequency-domain equalizer.
\end{remark}

\subsection{RAQR-Aware Equalization}
\label{subsec:raqr_equalization}
Assuming perfect knowledge of the nonzero composite subcarrier gain
$d_n=\sqrt{P_{t,n}}g(\delta_s^n)h_n$, the estimated symbol based on the RAQR-aware
zero-forcing (ZF) equalizer is 
\begin{equation}
\hat s_{{\rm ZF},n}
=\frac{y_n}{d_n}
=s_n+\frac{\hat w_n}{d_n}.
\label{eq:scalar_equalizers}
\end{equation}
Accordingly, the output SNR is given by 
\begin{equation}
\gamma_{{\rm ZF},n}
=\frac{P_{t,n}|g(\delta_s^n)|^2|h_n|^2}
{\sigma_n^2}.
\label{eq:zf_output_snr}
\end{equation} 
Then,
the achievable rate is expressed as
\begin{equation}
R_{\rm ZF}
=\Delta f
\sum_{n=1}^{N_c}
\log_2\left(1+\gamma_{{\rm ZF},n}\right).
\label{eq:zf_achievable_rate}
\end{equation}

\section{Bandwidth and Conversion-Gain Analysis}
\label{sec:gain_bandwidth}
This section characterizes the effects of the LO and AUX fields on the one-sided response and examines its tradeoff with bandwidth.

\subsection{Response and Performance Metrics}
\label{subsec:gb_response}
 
To characterize the receiver bandwidth, we normalize the complex response $H(\delta_s)=H_+(\delta_s)-H_-^*(\delta_s)$ by its zero-detuning value $H(0)\neq0$. At each fixed LO/AUX operating point, the normalized response and the corresponding \textcolor{black}{LO-referenced} bandwidth are defined as
\begin{equation}
R(\delta_s)=\frac{|g(\delta_s)|^2}{|g(0)|^2}\triangleq\frac{|H(\delta_s)|^2}{|H(0)|^2},
\label{eq:gb_R_def}
\end{equation}
\begin{equation}
W_+\triangleq\frac{1}{2\pi}\min\big\{x>0:\ R(-x)=1/2\big\}.
\label{eq:gb_wplus_def}
\end{equation}
Here, $|H(0)|$ characterizes the center atomic-response
amplitude, whereas the actual conversion coefficient satisfies
$|g(0)|=|g_{\rm c}||H(0)|$.
The readout prefactor $g_{\rm c}$ is independent of
$\delta_s$ but varies with the LO and AUX driving.
\textcolor{black}{Note that this definition imposes a lower threshold relative to
the center response rather than a bound on in-band ripple.
For nonmonotonic responses with off-center peaks, the bandwidth 
does not guarantee a ripple of at most $3$~dB and generally differs
from the peak-referenced 3-dB bandwidth.}

Owing to the analytical intractability of $R(\delta_s)$, analyzing the normalized response is nontrivial. To tackle this issue, we assume that the probe, coupling, and LO fields are perfectly resonant, namely $\Delta_p=\Delta_c=\Delta_L=0$. Substituting these conditions into~\eqref{D_3}--\eqref{D_5}, we have $D_3=\xi_A$, $D_4=-\xi_A$, and $D_5=\Delta_A$. Then, $Z_k(\delta_s)$ is simplified to $Z_k(\delta_s)=\zeta_k-\delta_s$ for $k\in\{2,3,4,5\}$, where
\begin{equation}
\begin{split}
&\zeta_2={\rm j}\eta_2,\qquad \zeta_3=\xi_A+{\rm j}\eta_r,\\
&\zeta_4=-\xi_A+{\rm j}\eta_r,\qquad \zeta_5=\Delta_A+{\rm j}\eta_r.
\end{split}
\label{eq:gb_zeta}
\end{equation}
The imaginary parts of these complex detunings are the decay rates of the coherences, whereas the real parts of $\zeta_3$ and $\zeta_4$ are the AC Stark shifts that the AUX field induces on $|3\rangle$ and $|4\rangle$.
Then,  these complex detunings and the field couplings are collected into the coherence matrix, which is given by
\begin{equation}
\mathbf M=\begin{bmatrix}\zeta_2&\Omega_c/2&0&0\\\Omega_c/2&\zeta_3&\Omega_L/2&\Omega_A/2\\0&\Omega_L/2&\zeta_4&0\\0&\Omega_A/2&0&\zeta_5\end{bmatrix}.
\label{eq:gb_M}
\end{equation}
Expanding $\mathcal Q(\delta_s)\triangleq\det(\mathbf M-\delta_s\mathbf I)$ along the row of $|3\rangle$, we have
\begin{equation}
 \mathcal{Q}  (\delta_s)=F(\delta_s)Z_4(\delta_s)Z_5(\delta_s).
\end{equation}
Substituting $F(\delta_s)=\frac{\mathcal Q(\delta_s)}{Z_4(\delta_s)Z_5(\delta_s)}$ into~\eqref{eq:rho21_plus_closed_main} and~\eqref{eq:rho21_minus_closed_main}, the response at perfect resonance is expressed as
\begin{equation}
H_r(\delta_s)=\frac{C_0\,\zeta_4\zeta_5(\zeta_5-\delta_s)}{\mathcal Q(0)\,\mathcal Q(\delta_s)}-\frac{C_0\,\zeta_5^*(\zeta_4^*+\delta_s)(\zeta_5^*+\delta_s)}{\mathcal Q^*(0)\,\mathcal Q^*(-\delta_s)},
\label{eq:gb_H_res}
\end{equation}
where $C_0=\frac{ \Omega_p\Omega_c^2\Omega_L}{32}$. Although~\eqref{eq:gb_H_res} is explicit, the roles of LO and AUX
driving are not apparent. To clarify these roles, we
approximate the resonance locations by exploiting two separated
frequency scales. The complex detunings, comprising decay rates
and AUX-induced shifts, govern the leading-order linewidths and
frequency shifts, whereas the coupling and LO fields set the
dominant splitting through the dressing scale
$\overline{\Omega}\triangleq\sqrt{\Omega_L^2+\Omega_c^2}$.

\begin{table*}[t]
\centering
\caption{Baseline Parameter Settings}
\label{tab:simulation_parameters}

\footnotesize
\renewcommand{\arraystretch}{1.28}
\setlength{\extrarowheight}{1pt}
\setlength{\tabcolsep}{2pt}

\begin{minipage}[t]{0.325\textwidth}
\vspace{0pt}
\centering
\rowcolors{2}{gray!4}{white}
\begin{tabularx}{\linewidth}
{@{}Y@{\hspace{2pt}}c@{\hspace{2pt}}c@{}}
\rowcolor{tableblue}
\multicolumn{3}{c}{\textbf{Atomic Driving and Relaxation}}\\
\toprule
\textbf{Parameter} & \textbf{Value} & \textbf{Unit}\\
\midrule
Probe Rabi frequency \(\Omega_p/(2\pi)\)
& \textcolor{black}{\(0.85\)} & MHz\\
Coupling Rabi frequency \(\frac{\Omega_c}{2\pi}\)
&  \textcolor{black}{\(15\)} & MHz\\
LO Rabi frequency \(\Omega_L/(2\pi)\)
&  \textcolor{black}{\(7.5\)} & MHz\\
AUX Rabi frequency \(\Omega_A/(2\pi)\)
& \textcolor{black}{\(150\)} & MHz\\
Probe detuning \(\Delta_p/(2\pi)\)
& \(0\) & MHz\\
Coupling detuning \(\Delta_c/(2\pi)\)
& \(0\) & MHz\\
LO detuning \(\Delta_L/(2\pi)\)
& \(0\) & MHz\\
AUX detuning \(\Delta_A/(2\pi)\)
& \textcolor{black}{\(1697\)} & MHz\\
Off-resonant detuning \(\Delta'/(2\pi)\)
& \textcolor{black}{\(3214\)} & MHz\\
\textcolor{black}{Effective dephasing rate} \(\gamma/(2\pi)\)
& \textcolor{black}{\(0.352\)} & MHz\\
Decay rate \(\gamma_2/(2\pi)\)
& \(6.07\) & MHz\\
Decay rate \(\gamma_3,\gamma_4,\gamma_5/(2\pi)\)
& \(2.2,2.2,2.2\) & kHz\\
\bottomrule
\end{tabularx}
\end{minipage}
\hfill
\begin{minipage}[t]{0.325\textwidth}
\vspace{0pt}
\centering
\rowcolors{2}{gray!4}{white}
\begin{tabularx}{\linewidth}
{@{}Y@{\hspace{2pt}}c@{\hspace{2pt}}c@{}}
\rowcolor{tableblue}
\multicolumn{3}{c}{\textbf{Optical and Cell Parameters}}\\
\toprule
\textbf{Parameter} & \textbf{Value} & \textbf{Unit}\\
\midrule
Probe dipole \(\lvert\mu_{12}\rvert\)
& \(2.989\) & \(ea_0\)\\
Coupling dipole \(\lvert\mu_{23}\rvert\)
& \(0.01527\) & \(ea_0\)\\
LO/signal dipole \(\lvert\mu_{34}\rvert\)
& \(2117.49\) & \(ea_0\)\\
AUX dipole \(\lvert\mu_{35}\rvert\)
& \(2176.08\) & \(ea_0\)\\
Probe wavelength \(\lambda_p\)
& \(780.241\) & nm\\
Cell length \(L_{\rm cell}\)
& \textcolor{black}{\(6.5\)} & cm\\
Atomic density \(N_0\)
& \textcolor{black}{\(5.89\times10^{14}\)} & \(\mathrm{m}^{-3}\)\\
Probe power \(P_{\rm in}\)
& \textcolor{black}{\(252\)} & \(\mathrm{nW}\)\\
Coupling power \(P_c\)
&\textcolor{black}{3}  & W\\
Beam diameter \(w_p=w_c\)
& \textcolor{black}{0.7} & mm\\
Reference optical power \(P_r\)
& \(30\) & mW\\
PD responsivity \(\mathcal R_{\rm pd}\)
& \(0.5494\) & A/W\\
\bottomrule
\end{tabularx}
\end{minipage}
\hfill
\begin{minipage}[t]{0.325\textwidth}
\vspace{0pt}
\centering
\rowcolors{2}{gray!4}{white}
\begin{tabularx}{\linewidth}
{@{}Y@{\hspace{2pt}}c@{\hspace{2pt}}c@{}}
\rowcolor{tableblue}
\multicolumn{3}{c}{\textbf{Readout and Communication}}\\
\toprule
\textbf{Parameter} & \textbf{Value} & \textbf{Unit}\\
\midrule
LO carrier frequency \(f_{\rm LO}\)
& \(\approx16.03\) & GHz\\
Electrical power gain \(G_{\rm A}\)
& \(30\) & dB\\
Load resistance \(R_L\)
& \(1\) & \(\Omega\)\\
Number of subcarriers \(N_c\)
& \(1024\) & --\\
Subcarrier spacing \(\Delta f\)
& \textcolor{black}{\(10\)} & kHz\\
Allocated bandwidth \(N_c\Delta f\)
& \textcolor{black}{\(10.24\)} & MHz\\
Total transmit power \(P_t\)
& \(23\) & dBm\\
Transmit antenna gain \(G_t\)
& \(0\) & dBi\\
Propagation distance \(d_{\rm dis}\)
& \(1500\) & m\\
Noise temperature \(T_{\rm pd}\)
& \textcolor{black}{\(100\)} & K\\
RF receiver noise figure
& \(7\) & dB\\
Effective RF receive gain
& \(3.951\) & dBi\\
\bottomrule
\end{tabularx}
\end{minipage}

\end{table*}

\subsection{Effects of a Far-Detuned AUX Field}
\label{subsec:gb_aux}
We first simplify the response by eliminating the coherence associated with $|5\rangle$, and obtain
a frequency-dependent $3\times3$ representation. Since $|5\rangle$
couples only to $|3\rangle$, the Schur complement of its diagonal
entry in $\mathbf M-\delta_s\mathbf I_4$ yields
\begin{equation}
\mathcal Q(\delta_s)
=(\zeta_5-\delta_s)
\det\!\big(\hat{\mathbf M}(\delta_s)-\delta_s\mathbf I_3\big).
\label{eq:gb_schur}
\end{equation}
For $\zeta_5\neq\delta_s$, $\hat{\mathbf M}(\delta_s)$ is the
upper-left $3\times3$ block of $\mathbf M$ with $\zeta_3$ replaced by
$[\zeta_3-\frac{\Omega_A^2}{4(\zeta_5-\delta_s)}]$.
The factors
$\zeta_5(\zeta_5-\delta_s)$ and
$\zeta_5^*(\zeta_5^*+\delta_s)$ in the numerators are cancelled by factorization, and thus the AUX influence is retained through the frequency-dependent correction to the complex
detuning of $|3\rangle$.

To approximate the AUX-induced correction as frequency-independent, we assume $|\Delta_A|\gg\max\{\Omega_A,|\delta_s|,\eta_r\}$
throughout the band of interest. Then,
with $\zeta_5-\delta_s\approx\Delta_A$, the shift of $|3\rangle$ reduces to the static dispersive shift $\frac{-\Omega_A^2}{4\Delta_A}$.
Accordingly, $\hat{\mathbf M}(\delta_s)$ is approximated by
\begin{equation}
\hat{\mathbf M}(\delta_s)\approx \widetilde{\mathbf M}=
\begin{bmatrix}
\zeta_2 & \Omega_c/2 & 0\\
\Omega_c/2 & \widetilde{\zeta}_3 & \Omega_L/2\\
0 & \Omega_L/2 & \zeta_4
\end{bmatrix},
\label{eq:gb_Mt}
\end{equation}
where $\widetilde{\zeta}_3=
\widetilde D_3+\mathrm{j}\eta_r$ and
\begin{equation}
\widetilde D_3=\xi_A-\frac{\Omega_A^2}{4\Delta_A}
=\frac{\Omega_A^2}{4}\left(\frac{\mu_{34}^2}{\mu_{35}^2\Delta'}-\frac{1}{\Delta_A}\right).
\label{eq:gb_D3t}
\end{equation}
Therefore, we have $\mathcal Q(\delta_s)\approx(\zeta_5-\delta_s)\mathcal P(\delta_s)$, where \(\mathcal P(\delta_s)\) is 
\begin{equation}
\begin{split}
\mathcal P(\delta_s)=&\det(\widetilde{\mathbf M}-\delta_s\mathbf I)\\
={}&(\zeta_2-\delta_s)(\widetilde\zeta_3-\delta_s)(\zeta_4-\delta_s)\\
-&(\zeta_2-\delta_s)\frac{\Omega_L^2}{4}-(\zeta_4-\delta_s)\frac{\Omega_c^2}{4}.
\end{split}
\label{eq:gb_P}
\end{equation}
Substituting this result into~\eqref{eq:gb_H_res}, we have 
\begin{equation}
H_r(\delta_s)\approx\frac{C_0\,\zeta_4}{\mathcal P(0)\,\mathcal P(\delta_s)}-\frac{C_0\,(\zeta_4^*+\delta_s)}{\mathcal P^*(0)\,\mathcal P^*(-\delta_s)}.
\label{eq:gb_H_reduced}
\end{equation}

\begin{remark}[Effects of AUX MW]\label{AUXMWeffects}
    The far-detuned AUX field influences the receiver primarily through dispersive dressing rather than resonant population transfer. The admixture of \(|5\rangle\) into \(|3\rangle\) scales as \(\frac{\Omega^2_A}{4\Delta^2_A}\) and remains small under the far-detuning assumption. Its dominant effect is to introduce static level shifts. Specifically, \(|3\rangle\) experiences the effective shift \(\widetilde D_3\), which combines the AC Stark shift \(\xi_A\) from the off-resonant \(|3\rangle\leftrightarrow|4\rangle\) coupling and the dispersive shift \(\frac{-\Omega_A^2}{4\Delta_A}\) induced by \(|5\rangle\), while \(|4\rangle\) is shifted by \(-\xi_A\). The five-level receiver can therefore be interpreted as an effective four-level system with tunable Rydberg-level detunings. As shown below, these shifts play distinct roles. The shift \(\widetilde D_3\) mainly controls the position of the Autler-Townes doublet, whereas \(\xi_A\) primarily modifies the conversion gain.
\end{remark}

\subsection{Autler--Townes Doublet}
\label{subsec:gb_doublet}
According to~\eqref{eq:gb_P}, $\mathcal P(\delta_s)$ is the
characteristic polynomial of $\widetilde{\mathbf M}$. Therefore, the resonances of the two terms in~\eqref{eq:gb_H_reduced} are
associated with its eigenvalues and their mirror images about
the LO frequency, respectively. Physically, the coupling and LO
fields form three dressed states from $|2\rangle$, $|3\rangle$,
and $|4\rangle$. In the lossless, zero-detuning limit,
their eigenvalues are $0$ and $\pm\overline{\Omega}/2$.
The dark state contains no $|3\rangle$ component
because the two coupling pathways interfere destructively,
whereas the bright states form an
Autler-Townes doublet with splitting $\overline{\Omega}$.

Decay rates broaden the dressed-state resonances, whereas AUX-induced detunings shift their frequencies. When these diagonal perturbations are small relative to the dressed-state splitting, first-order perturbation theory yields eigenvalue corrections weighted by the corresponding unperturbed dressed-state populations. Accordingly, $\mathcal P(\delta_s)$ is approximated by
\begin{equation}
\mathcal P(\delta_s)\approx
(\lambda_0-\delta_s)
\left[
(\delta_s-\beta-\mathrm{j}\alpha)^2
-\frac{\overline{\Omega}^2}{4}
\right],
\label{eq:gb_factor}
\end{equation}
where
$\lambda_0=\frac{\zeta_2\Omega_L^2+\zeta_4\Omega_c^2}
{\overline{\Omega}^2}$
denotes the approximate dark-mode eigenvalue, while
$\beta=\frac{\xi_A\Omega_c^2}{2\overline{\Omega}^2}
-\frac{\Omega_A^2}{8\Delta_A}$
and
$\alpha=\eta_r+\frac{(\eta_2-\eta_r)\Omega_c^2}
{2\overline{\Omega}^2}$
represent the frequency shift and decay rate of the bright modes,
respectively.
 
\begin{remark}[Effects of LO and AUX]\label{rem:gb_doublet}
Physically, increasing $\Omega_L$ enlarges the
doublet splitting $\overline{\Omega}$ and reduces its linewidth
by decreasing the contribution of the short-lived state $|2\rangle$.
In contrast, AUX driving shifts the bright-mode doublet through
$\beta$ and perturbs the dark resonance through $\xi_A$.  With AUX driving, the dark resonance shifts to $\Re\{\lambda_0\}\approx-\frac{\xi_A\Omega_c^2}{\overline{\Omega}^2}<0$ for $\xi_A>0$, contributing to an asymmetric atomic response about the LO. Under the considered operating conditions, this asymmetry favors negative signal–LO offsets, motivating the allocation of data subcarriers below the LO frequency.
\end{remark}

\subsection{Impacts on Gain and Bandwidth}
The dark resonance primarily determines the center atomic gain. By setting $\delta_s=0$ in~\eqref{eq:gb_H_reduced}, we have
\begin{equation}
H_r(0)=2\mathrm{j}C_0
\Im\left\{\frac{\zeta_4}{\mathcal P^2(0)}\right\}.
\label{eq:gb_H0}
\end{equation}
Since
$\mathcal P(0)=\frac{-\overline{\Omega}^2\lambda_0}{4}
+\zeta_2\widetilde\zeta_3\zeta_4$,
the EIT condition
$|\widetilde\zeta_3|\ll\frac{\Omega_c^2}{4\eta_2}$
allows us to approximate
$\mathcal P(0)\approx\frac{-\overline{\Omega}^2\lambda_0}{4}$.
Substituting this equation into~\eqref{eq:gb_H0}, \(|H_r(0)|\) is approximated by
\begin{equation}
|H_r(0)|\approx
\frac{\Omega_p\Omega_c^2\Omega_L
\left[
\eta_rS^2+\xi_A^2\Omega_c^2
(2S-\eta_r\Omega_c^2)
\right]}
{(S^2+\xi_A^2\Omega_c^4)^2},
\label{eq:gb_gain}
\end{equation}
where $S\triangleq\eta_2\Omega_L^2+\eta_r\Omega_c^2$. 
Under AUX driving, the center atomic gain
in~\eqref{eq:gb_gain} depends jointly on the LO Rabi frequency
$\Omega_L$ and the AUX-induced shift $\xi_A$.
Increasing $\Omega_L$ enhances RF-induced modulation but also
increases $S$, which may suppress the center response.
Meanwhile, AUX driving modifies both the numerator and denominator
of~\eqref{eq:gb_gain} through $\xi_A$.
Consequently, AUX driving can enhance the center gain when its
positive contribution outweighs the detuning-induced loss.
Specifically, for a fixed LO operating point, the approximate
gain exceeds its AUX-free value when
\begin{equation}
S^2(2S-3\eta_r\Omega_c^2)
>\eta_r\xi_A^2\Omega_c^6.
\label{eq:gb_aux_gain_condition}
\end{equation}

The bright states primarily determine the side peaks.
Near $\delta_s\approx-(\frac{\overline{\Omega}}{2}+\beta)$, the second term
of~\eqref{eq:gb_H_reduced} dominates when
$|\zeta_4|,|\lambda_0|\ll\frac{\overline{\Omega}}{2}$.
The corresponding peak magnitude is approximated by
$|H_r|_{\rm pk}\approx
\frac{4C_0}{\overline{\Omega}^3\alpha|\lambda_0|}$, and
the peak-to-center ratio is given by
\begin{equation}
\frac{|H_r|_{\rm pk}}{|H_r(0)|}
\approx
\frac{\overline{\Omega}}
{8\alpha|\lambda_0|
\left|\Im\{\zeta_4/\lambda_0^2\}\right|}.
\label{eq:gb_ripple}
\end{equation}
Without AUX driving, this ratio reduces to
$\frac{\overline{\Omega}\Im\{\lambda_0\}}{8\alpha\eta_r}$
and increases with $\Omega_L$ for $\eta_2>\eta_r$.
In contrast, AUX driving increases $|\lambda_0|$ and
reduces the side peak by the factor
$(1+\xi_A^2\Omega_c^4/S^2)^{-1/2}$ with a fixed LO.

\begin{remark}[Effects of LO and AUX driving on the gain-bandwidth tradeoff]\label{gb_su}
Increasing $\Omega_L$ enlarges the dressed-state splitting
$\overline{\Omega}$ and can broaden the continuous \textcolor{black}{LO-referenced} bandwidth
by shifting the bright-mode resonances outward. However, the
center atomic gain does not increase monotonically and
eventually decreases in the LO-dominated regime.
In contrast, AUX driving primarily adjusts the dressed-state
detunings and can enhance the center gain
under~\eqref{eq:gb_aux_gain_condition}. Nevertheless, under the
considered operating conditions, the increased center response
and suppressed outer-band response can narrow the normalized
\textcolor{black}{LO-referenced} bandwidth. Therefore, LO and AUX driving exhibit distinct
effects on the gain-bandwidth tradeoff to balance sensitivity and bandwidth.
\end{remark}

\subsection{One-Sided \textcolor{black}{LO-referenced} Bandwidth and Gain--Bandwidth Tradeoff}
\label{subsec:gb_bandwidth}

To characterize the gain-bandwidth tradeoff analytically,
we first consider the AUX-free receiver, for which
\eqref{eq:gb_H_reduced} is simplified as
\begin{equation}
H_r^{\rm AF}(\delta_s)=
\frac{C_0(2\mathrm{j}\eta_r-\delta_s)}
{\mathcal P(0)\mathcal P(\delta_s)}.
\label{eq:gb_H_AF}
\end{equation}
The corresponding center atomic gain is
$|H_r^{\rm AF}(0)|=\frac{2C_0\eta_r}{|\mathcal P(0)|^2}$.
Based on this response, we establish the following
gain-bandwidth inequality.

\begin{proposition}[Gain-bandwidth inequality]
\label{prop:gb_bandwidth}
For the AUX-free receiver, let
$\mathcal B=\frac{\overline{\Omega}^2}{4}
+2\eta_2\eta_r+\eta_r^2$.
The one-sided \textcolor{black}{LO-referenced} bandwidth satisfies
\begin{equation}
(2\pi W_+)^2
\le
\mathcal B+
\sqrt{
\frac{C_0}{\eta_r|H_r^{\rm AF}(0)|}
\left(1+\frac{4\eta_r^2}{\mathcal B}\right)
}.
\label{eq:gb_one_sided_bandwidth}
\end{equation}
\end{proposition}

\begin{IEEEproof}
For the AUX-free response, direct expansion gives \(|\mathcal P(\delta_s)|^2 =\delta_s^2(\delta_s^2-\mathcal B)^2+ \left[
(\eta_2+2\eta_r)\delta_s^2
-|\mathcal P(0)|
\right]^2\).
If $(2\pi W_+)^2\leq\mathcal B$,
the claimed bound holds directly.
Otherwise, for $\delta_s^2>\mathcal B$,
discarding the nonnegative second term yields \(R(\delta_s)\leq
\frac{|\mathcal P(0)|^2}
{(\delta_s^2-\mathcal B)^2}
\left(
\frac{1}{\delta_s^2}
+\frac{1}{4\eta_r^2}
\right)\).
Evaluating this inequality at the half-power boundary
$\delta_s=2\pi W_+$, where $R(2\pi W_+)=1/2$,
and using
$|H_r^{\rm AF}(0)|
=2C_0\eta_r/|\mathcal P(0)|^2$
establishes~\eqref{eq:gb_one_sided_bandwidth}.
\end{IEEEproof}

Proposition~\ref{prop:gb_bandwidth} constrains the achievable
bandwidth relative to the center atomic-response amplitude.
For fixed $\mathcal B$ and $C_0$, increasing the center gain
tightens the bandwidth bound. However, LO driving changes both
quantities, and thus the actual tradeoff depends on the operating
regime. As a result, we examine the strong-LO limit.

\begin{corollary}[Strong-LO gain--bandwidth tradeoff]
\label{cor:gb_tradeoff}
For the AUX-free receiver with
$\eta_2>\sqrt{2}\eta_r$,
as $\Omega_L\to\infty$ with all other parameters fixed,
the one-sided bandwidth satisfies
\begin{equation}
|H_r^{\rm AF}(0)|^2(2\pi W_+)^6
\to
\frac{\Omega_p^2\Omega_c^4}
{128\sqrt{2}\eta_2\eta_r}
\left(
1+\frac{\sqrt{2}\eta_r}{\eta_2}
\right)^3.
\label{eq:gb_atomic_tradeoff}
\end{equation}
\end{corollary}

\begin{IEEEproof}
In the strong-LO limit,
$|\mathcal P(0)|\sim\frac{\eta_2\Omega_L^2}{4}$,
giving $|H_r^{\rm AF}(0)|\sim \frac{\Omega_p\Omega^2_c\eta_r}{\eta^2_2\Omega^3_L}$.
Meanwhile, the \textcolor{black}{LO-referenced} bandwidth satisfies
$2\pi W_+\sim \frac{\Omega_L}{2}\sqrt{1+\frac{\eta_2}{\sqrt{2}\eta_r}}$.
Combining their leading-order expressions yields
\eqref{eq:gb_atomic_tradeoff}.
\end{IEEEproof}

Corollary~\ref{cor:gb_tradeoff} reveals that the center atomic-response power scales as $W_+^{-6}$ in the AUX-free strong-LO regime. Consequently, doubling the one-sided bandwidth incurs an asymptotic power loss of approximately $18.1$~dB.

With AUX driving, the dispersive shifts modify both the
center response and the bright-mode resonances.
Under the preceding far-detuned and dominant-response
approximations, we consider operating points where the
outer half-power crossing is also the first half-power crossing below the LO. Then, the \textcolor{black}{LO-referenced} bandwidth 
approximately satisfies
\begin{equation}
2\pi W_+\approx\sqrt{\frac{\overline{\Omega}^2}{4}-\alpha^2+\sqrt{\mathcal{K}^2-\overline{\Omega}^2\alpha^2}}+\beta,
\label{eq:gb_W_aux}
\end{equation}
where \(\mathcal{K}= \frac{\sqrt2|C_0|}{|\mathcal P(0)|\,|H_r(0)|} \).
This relation reveals that AUX driving can enhance the center
gain by adjusting the dark-state resonance, while the induced
frequency shift $\beta$ modifies the band edge.
Under the considered detuning conditions, improved the center gain
may therefore come at the expense of a narrower continuous
\textcolor{black}{LO-referenced} bandwidth, highlighting the gain-bandwidth tradeoff.

\begin{figure}[t]
    \centering
    \includegraphics[width=1\linewidth]{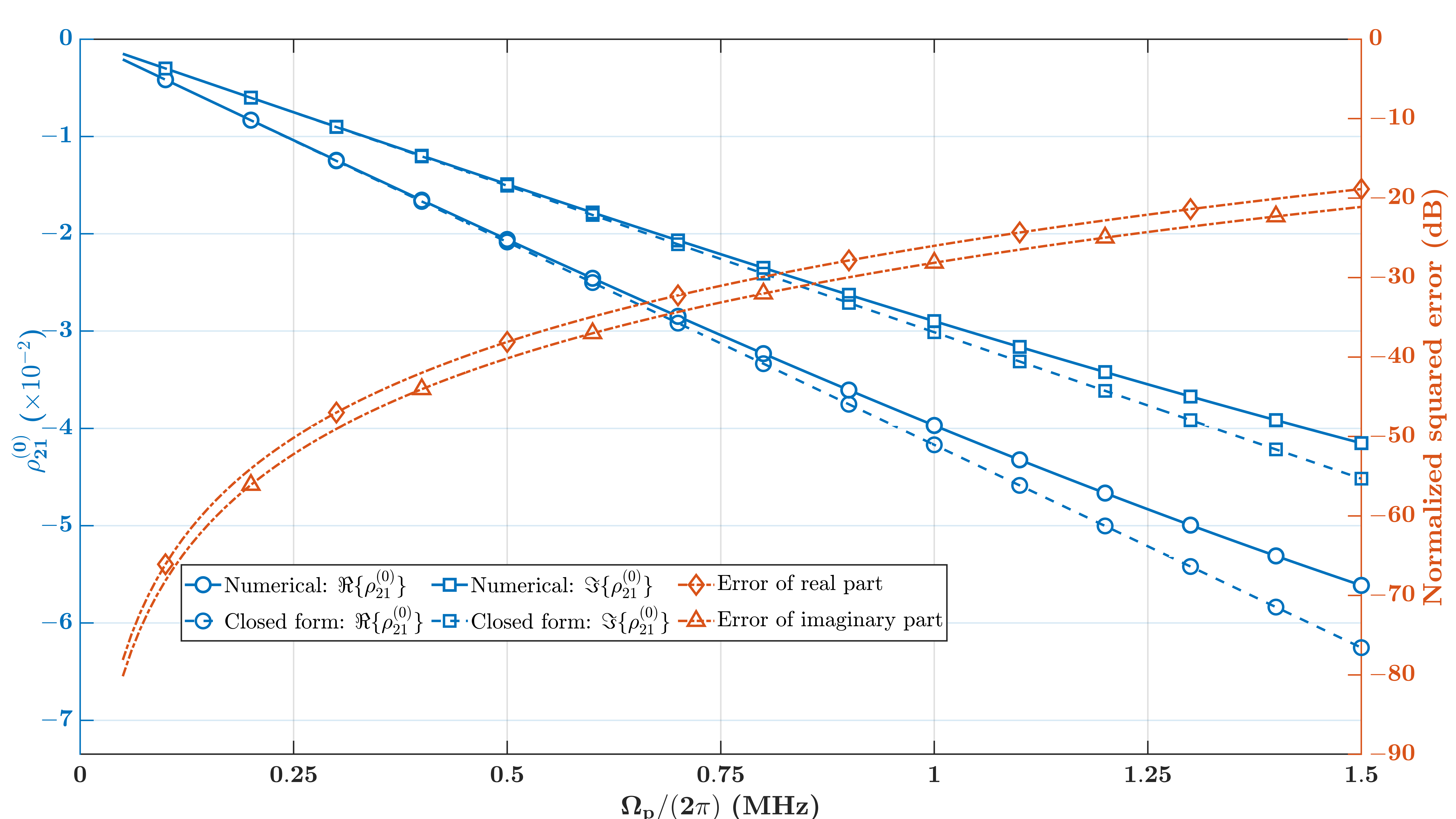}
    \caption{\textcolor{black}{\(\rho_{21}^{(0)}\) under various probe beams.}}
    \label{rho210}
\end{figure}

\begin{figure*}[t]
	\centering
	 \includegraphics[width=.9\linewidth]{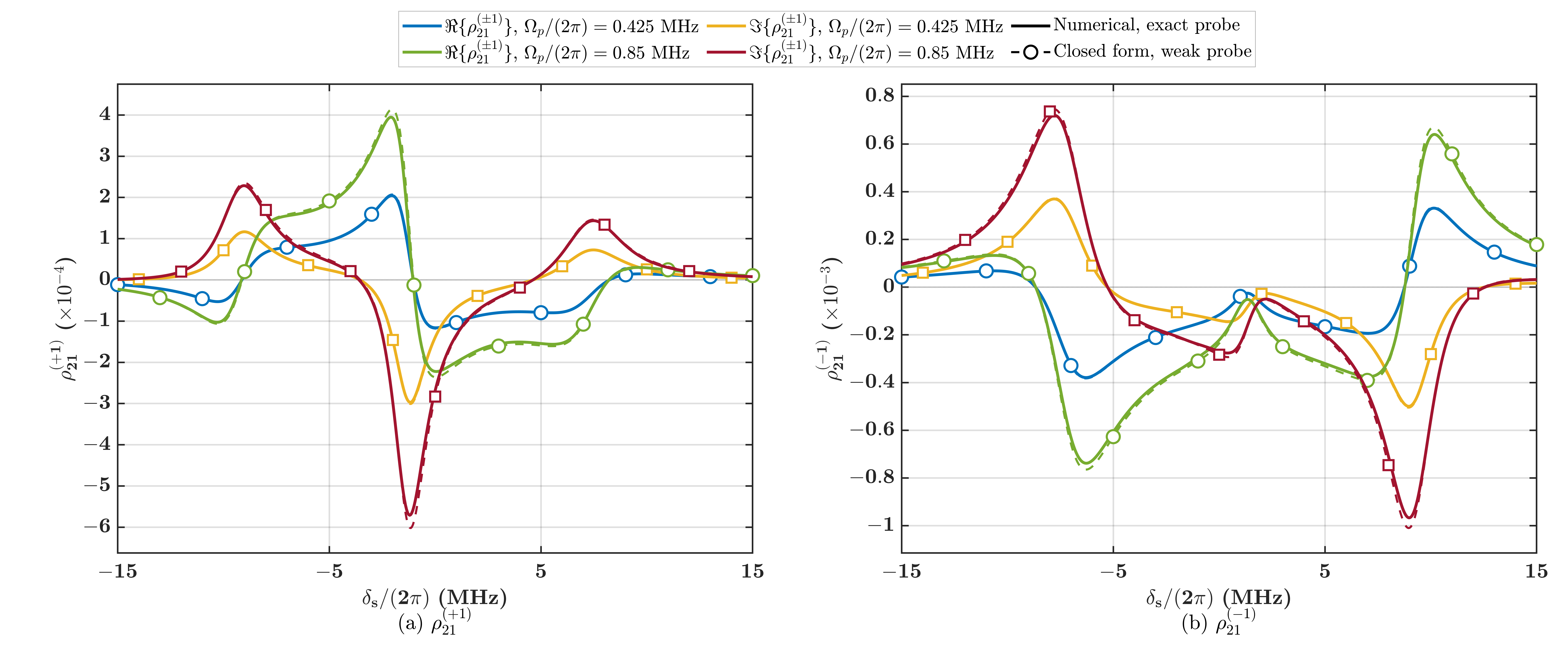}
	\caption{\textcolor{black}{\(\rho_{21}^{(+)}\) and \(\rho_{21}^{(-)}\) under different frequency offsets between the LO and the signal.}}
    \label{rho21+-}
\end{figure*}
\begin{figure}[t]
    \centering
    \includegraphics[width=.9\linewidth]{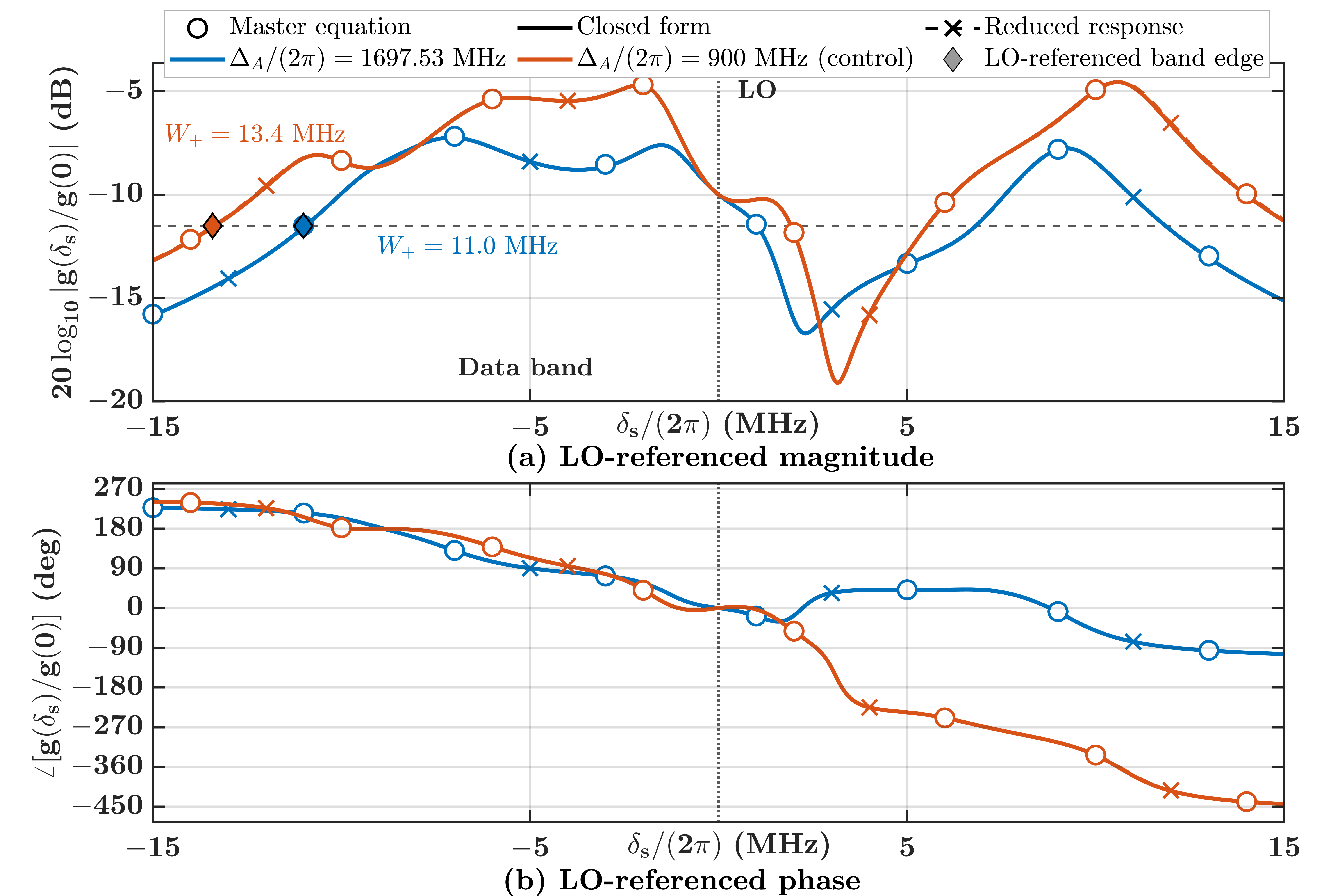}
    \caption{\textcolor{black}{Magnitude and phase response of RAQR under various detunings between the LO and user's signal.}}
    \label{raqr_frequency_response}
\end{figure}
\begin{figure}[t]
    \centering
    \includegraphics[width=.9\linewidth]{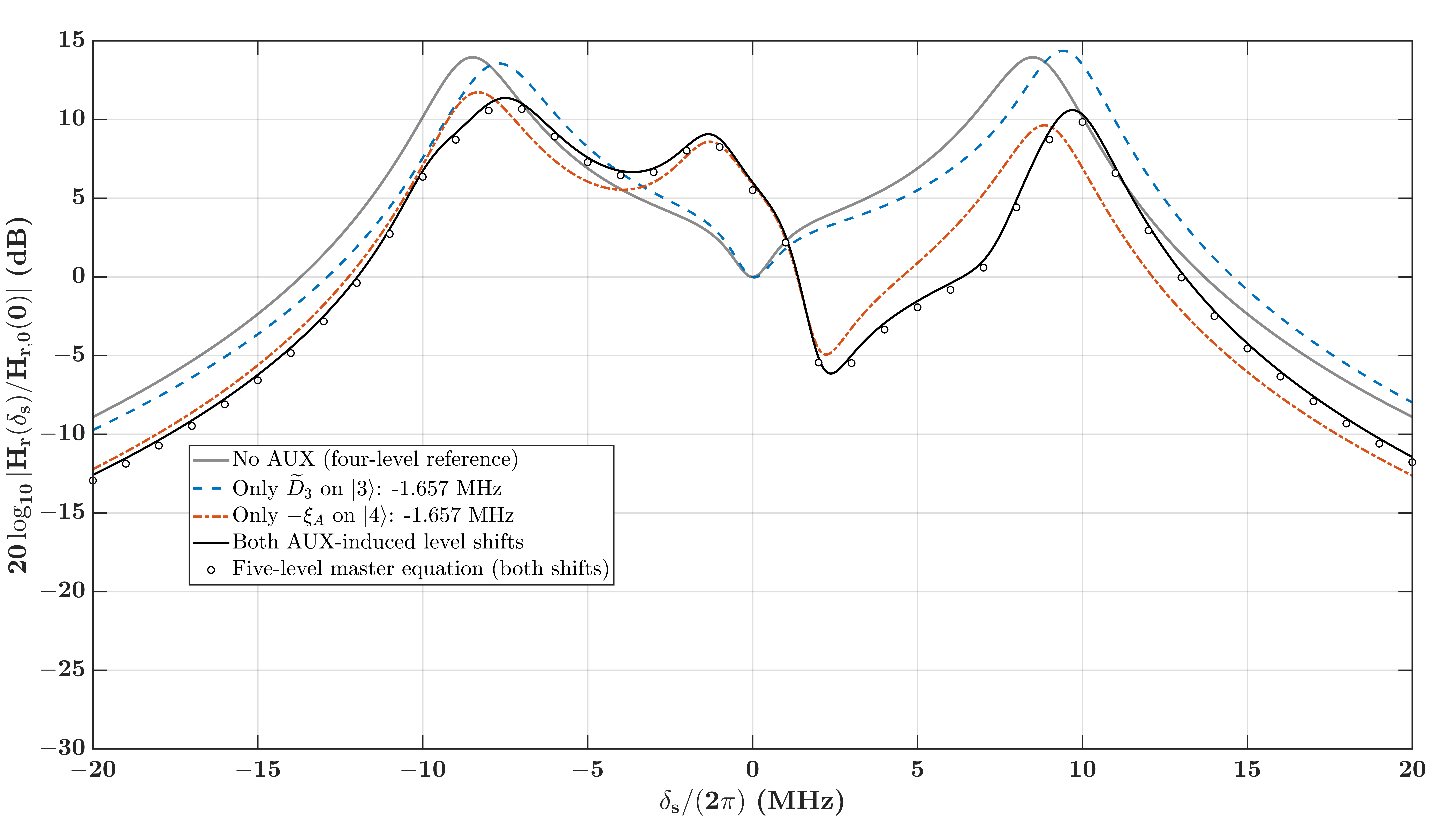}
    \caption{\textcolor{black}{Effects of AUX-induced level shifts on the atomic
response. The individual-shift curves retain only one shift contribution
to isolate its effect.}}
    \label{AXUshift}
\end{figure}

\section{Simulation Results}
\label{sec:simulation}
{This section evaluates the analytical model, the frequency response, and multicarrier equalization under the considered operating conditions.}

\subsection{Parameter Settings}
\label{subsec:parameter_settings}

We adopt the five-level \(^{87}\mathrm{Rb}\) scheme in~\cite{yan2026enhancinginstantaneousbandwidthrydberg}, with
\textcolor{black}{\(|1\rangle=|5S_{1/2},F=2\rangle\),
\(|2\rangle=|5P_{3/2},F'=3\rangle\),
\(|3\rangle=|51D_{5/2},m_J=5/2\rangle\),
\(|4\rangle=|52P_{3/2},m_J=3/2\rangle\), and
\(|5\rangle=|50F_{7/2},m_J=7/2\rangle\)}. Unlike the thermal-vapor configuration, we consider a laser-cooled atomic ensemble at $T_a =100~\mu \rm K$, where residual Doppler broadening is assumed negligible, allowing the stationary-atom approximation without Doppler averaging.
 Unless otherwise stated, the parameters are listed in Table~\ref{tab:simulation_parameters}. Furthermore, the phase of LO is set to \(\phi_{\rm LO}=0\).

\subsection{Closed-Form Expression Validation}
We first validate the closed-form expressions derived under the weak-probe approximation. Fig.~\ref{rho210} compares the real and imaginary parts of the derived steady-state response with numerical results obtained from the five-level Lindblad master equation for different probe Rabi frequencies. \textcolor{black}{For the considered parameter setting, particularly \(\Omega_c/(2\pi)=15~\mathrm{MHz}\), the agreement in Fig.~\ref{rho210} supports the weak-probe approximation for \(\Omega_p/(2\pi)<0.9~\mathrm{MHz}\).} As the probe Rabi frequency increases, the discrepancy
becomes more pronounced because population redistribution
and saturation weaken the assumptions
$\rho_{11}\approx1$ and $\rho_{ii}\approx0$ for $i=2,3,4,5$.

We further validate the first-order Floquet components \(\rho^{(+)}_{21}\) and \(\rho^{(-)}_{21}\) within the weak-probe and weak-signal regimes. Fig.~\ref{rho21+-} compares their real and imaginary parts with numerical results obtained from the five-level Lindblad master equation over different signal–LO detunings. We observe that the closed-form and numerical results remain in close agreement over the considered detuning range, demonstrating that the derived expressions accurately capture the frequency-dependent responses of both first-order Floquet components. Together with the steady-state validation, this agreement supports the accuracy of the closed-form expressions in characterizing the wideband atomic response.

\subsection{Atomic Response}

Fig.~\ref{raqr_frequency_response} illustrates the magnitude and phase responses across different signal-LO detunings. These frequency-dependent variations introduce subcarrier-dependent distortions even under a frequency-flat wireless channel, supporting Remark~\ref{remarkFDR} and motivating subcarrier-dependent equalization. It can be seen from the figure that the one-sided \textcolor{black}{LO-referenced} bandwidth reaches approximately $11~\mathrm{MHz}$. Moreover, the reduced model in~\eqref{eq:gb_H_reduced} closely matches the full response under sufficiently far-detuned AUX driving. Furthermore, Fig.~\ref{AXUshift} reveals the distinct roles of AUX-induced level shifts. The shift of $|3\rangle$ primarily displaces the bright-state doublet, whereas the shift of $|4\rangle$ modifies the dark-state resonance and can enhance the center conversion gain, consistent with Remark~\ref{AUXMWeffects}.

\subsection{Gain–Bandwidth Tradeoff}
\begin{figure}[t]
    \centering
    \includegraphics[width=0.9\linewidth]{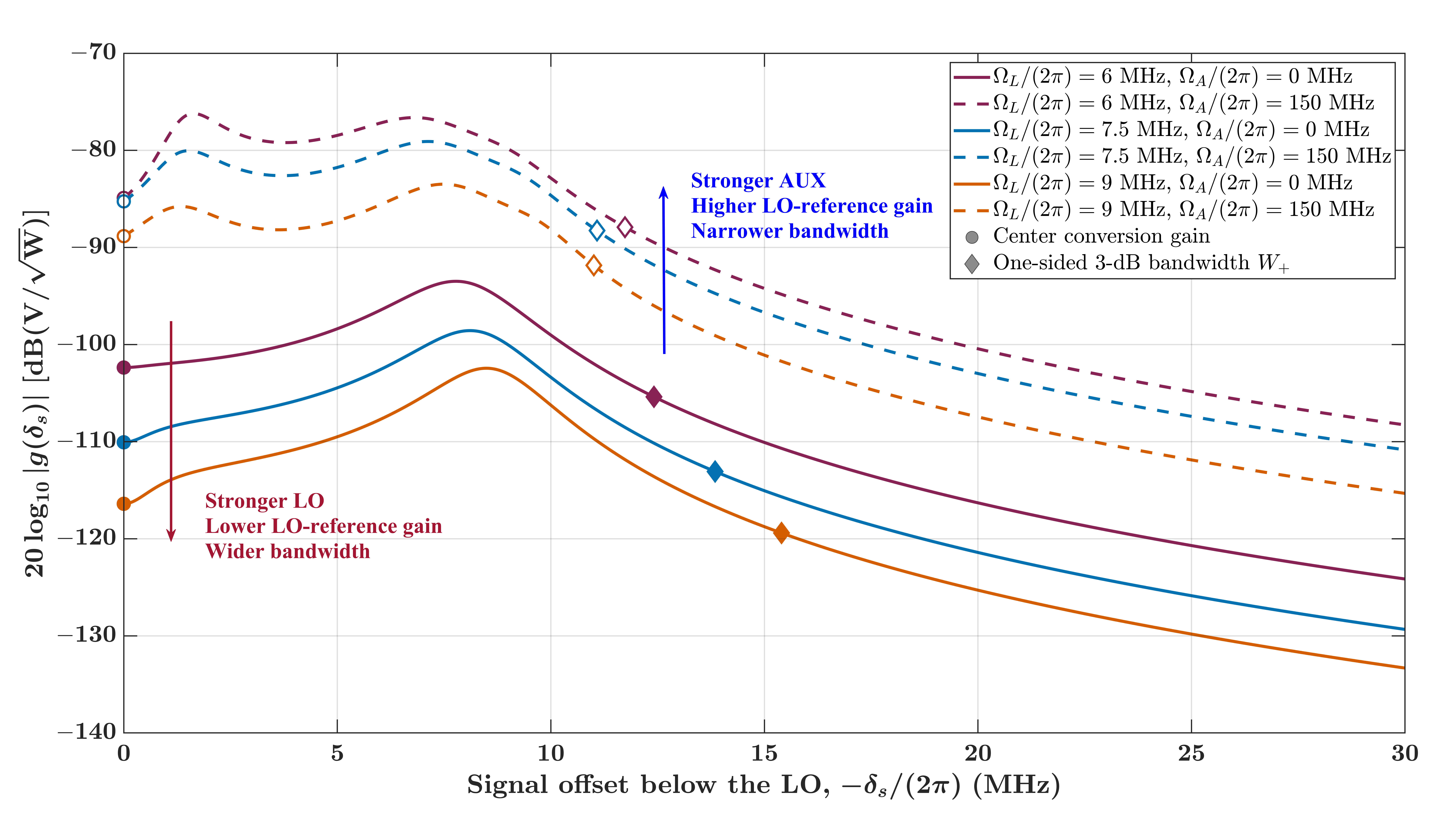}
    \caption{\textcolor{black}{Effects of AUX and LO on response.}}
    \label{AUXtradeoff}
\end{figure}

\begin{figure}[t]
    \centering
    \includegraphics[width=0.9\linewidth]{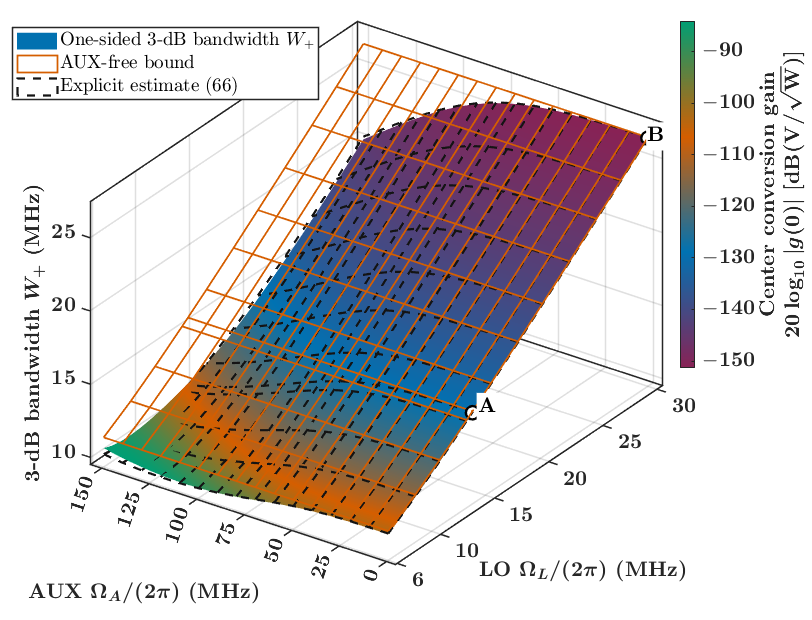}
    \caption{\textcolor{black}{Tradeoff between the center gain and bandwidth.}}
    \label{bound}
\end{figure}

Fig.~\ref{AUXtradeoff} illustrates the
effects of LO and AUX driving on the conversion gain and
one-sided \textcolor{black}{LO-referenced} bandwidth. Without AUX driving, increasing
$\Omega_L$ shifts the bright-state resonances outward and
narrows their linewidths by reducing the contribution of
the short-lived state $|2\rangle$. Meanwhile, the increased
$|2\rangle$ contribution to the dark state broadens the
central resonance and reduces the center gain. Consequently,
$W_+$ increases from $12.6$ to $15.7$~MHz at the cost of
approximately $13.5$~dB in center gain. \textcolor{black}{In contrast, far-detuned AUX driving modifies the center and outer-band responses through dispersive level shifts. Under the considered operating conditions, the center response increases more strongly than the outer-band response in relative terms, thereby reducing the normalized response $R(\delta_s)=|H(\delta_s)|^2/|H(0)|^2$ near the band edge. Consequently, the first half-power crossing shifts toward the LO, revealing that enhanced center conversion gain comes at the expense of a narrower LO-referenced bandwidth.} Together,
these results support Remarks~\ref{rem:gb_doublet} and
\ref{gb_su},
highlighting the distinct roles of LO and AUX driving
in shaping the gain-bandwidth tradeoff.

Fig.~\ref{bound} illustrates the \textcolor{black}{LO-referenced} bandwidth $W_+$ versus the LO and AUX Rabi frequencies, with color indicating the center conversion gain $20\log_{10}|g(0)|$. Increasing $\Omega_L$ broadens the bandwidth at the expense of center gain, whereas increasing $\Omega_A$ produces the opposite trend. The numerical bandwidth agrees with the analytical expression in~\eqref{eq:gb_W_aux}, while the AUX-free results satisfy the upper bound established in Proposition~\ref{prop:gb_bandwidth}. Moreover, the AUX-free operating points A and B agree well with the gain-bandwidth scaling predicted by Corollary~\ref{cor:gb_tradeoff}. These results confirm the analyzed tradeoff and provide guidance for LO/AUX parameter selection.

\subsection{Wideband Multicarrier Receiver}
\begin{figure*}
    \centering
       \includegraphics[width=1\linewidth]{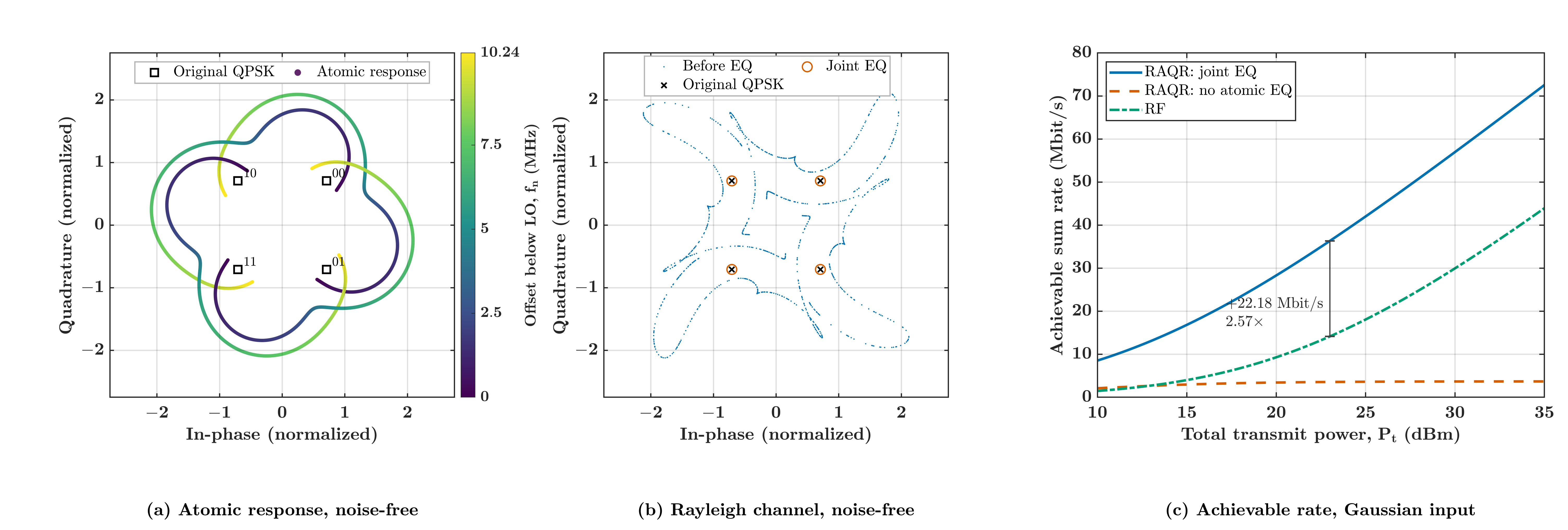}
    \caption{\textcolor{black}{Constellations of input signal, RAQR output with equalization, and achievable-rate performance.}}
    \label{constalation}
\end{figure*}
Fig.~\ref{constalation}(a) reveals subcarrier-dependent amplitude and phase distortions caused by the frequency-selective atomic response, even under a frequency-flat wireless channel. Fig.~\ref{constalation}(b) further illustrates the impact of multipath fading, with joint ZF equalization restoring the transmitted symbols in the noiseless case. With the presence of receiver noise, Fig.~\ref{constalation}(c) shows that frequency-flat atomic calibration leads to rate saturation under mismatched decoding, whereas RAQR-aware equalization compensates for the frequency-dependent atomic response. \textcolor{black}{As observed, at a transmit power of $23$~dBm, the RAQR achieves a data rate $2.57$ times that of the conventional RF benchmark, demonstrating the importance of RAQR-aware equalization and the potential of RAQRs for wideband multicarrier reception.}

\section{Conclusions}
This paper investigated continuous wideband OFDM reception using a five-level RAQR with a single microwave LO. The analysis revealed that the frequency-selective atomic response introduces subcarrier-dependent amplitude and phase distortions, highlighting the importance of RAQR-aware ZF equalization. Furthermore, LO and AUX driving exhibited distinct effects on the gain--bandwidth tradeoff. Under the considered operating conditions, stronger LO driving broadened the bandwidth at the expense of center response, whereas appropriate AUX driving enhanced the center response while narrowing the contiguous passband. In the AUX-free strong-LO regime, doubling the bandwidth incurred an approximately $18$-dB reduction in center-response power. Simulation results supported these findings and demonstrated a contiguous one-sided \textcolor{black}{LO-referenced} bandwidth exceeding $11$~MHz. These insights provide guidance for balancing bandwidth and conversion gain, highlighting the potential of RAQRs as wideband receiving front ends for future wireless communications.

\appendices
\section{Proof of the Weak-Probe Closed-Form Optical Coherence}
\label{app:weak_probe_closed_form}
Substituting the Hamiltonian in~\eqref{eq:H_full} into the
master equation~\eqref{mastereq}, we obtain the evolution
equations for the coherences $\rho_{k1}$, $k=2,3,4,5$.
Defining $\varphi=\delta_s t+\theta_s$, these equations
are given by
\begin{equation}
\begin{aligned}
\dot\rho_{21}
=&-\eta_2\rho_{21}
-\mathrm{j}\left[
\frac{\Omega_p}{2}(\rho_{11}-\rho_{22})
-\Delta_p\rho_{21}
+\frac{\Omega_c}{2}\rho_{31}
\right],\\
\dot\rho_{31}
=&-\eta_r\rho_{31}
-\mathrm{j}\Bigg[
\frac{\Omega_c}{2}\rho_{21}
-D_3\rho_{31}
+\frac{\Omega_L+\Omega_se^{-\mathrm{j}\varphi}}{2}
\rho_{41}\\
+&\frac{\Omega_A}{2}\rho_{51}
-\frac{\Omega_p}{2}\rho_{32}
\Bigg],\\
\dot\rho_{41}
=&-\eta_r\rho_{41}
-\mathrm{j}\left[
\frac{\Omega_L+\Omega_se^{\mathrm{j}\varphi}}{2}
\rho_{31}
-D_4\rho_{41}
-\frac{\Omega_p}{2}\rho_{42}
\right],\\
\dot\rho_{51}
=&-\eta_r\rho_{51}
-\mathrm{j}\left[
\frac{\Omega_A}{2}\rho_{31}
-D_5\rho_{51}
-\frac{\Omega_p}{2}\rho_{52}
\right].
\end{aligned}
\label{eq:A_exact}
\end{equation}
These equations are obtained without invoking the weak-probe
approximation. Since the repopulation matrix is diagonal, it
does not contribute to the off-diagonal coherence dynamics,
while the anticommutator term determines the coherence decay rates.

Under the weak-probe approximation, the atomic population remains
predominantly in the ground state, yielding $\rho_{11}\approx1$
and $\rho_{kk}\approx0$, $k=2,3,4,5$. Meanwhile, coherences
between excited states are neglected, i.e.,
$\rho_{k2}\approx0$, $k=3,4,5$. Consequently,
$\rho_{11}-\rho_{22}\approx1$, and~\eqref{eq:A_exact}
reduces to a closed set of linear equations for
$\rho_{21}$, $\rho_{31}$, $\rho_{41}$, and $\rho_{51}$,
while retaining the coupling, LO, and AUX fields
without assuming weak driving.

Then, with the weak-signal, i.e.,
$|\Omega_s|\ll|\Omega_L|$, the coherences are expanded as
\begin{equation}
\rho_{k1}(t)\approx
\rho_{k1}^{(0)}
+\rho_{k1}^{(+)}e^{\mathrm{j}\varphi}
+\rho_{k1}^{(-)}e^{-\mathrm{j}\varphi},
\end{equation}
where the terms of higher order $\mathcal O(\Omega_s^2)$ are neglected. Substituting the Floquet expansions into~\eqref{eq:A_exact}
and retaining terms up to first order in $\Omega_s$
yields three linear systems with the same coefficient structure.
Therefore, these systems can be expressed in a unified matrix form as
\begin{equation}
\mathbf A(\nu_q)\mathbf x^{(q)}=\mathbf b^{(q)},
\quad q\in\{0,+,-\},
\label{eq:A_linear_system}
\end{equation}
where $\nu_0=0$, $\nu_+=\delta_s$, and $\nu_-=-\delta_s$.
The coherence vector is defined as
\begin{equation}
\mathbf x^{(q)}
=
\left[
\rho_{21}^{(q)},
\rho_{31}^{(q)},
\rho_{41}^{(q)},
\rho_{51}^{(q)}
\right]^{\rm T}.
\end{equation}
The coefficient matrix is given by
\begin{equation}
\mathbf A(\nu)=
\begin{bmatrix}
Z_2(\nu)&-\Omega_c/2&0&0\\
-\Omega_c/2&Z_3(\nu)&-\Omega_L/2&-\Omega_A/2\\
0&-\Omega_L/2&Z_4(\nu)&0\\
0&-\Omega_A/2&0&Z_5(\nu)
\end{bmatrix},
\label{eq:A_matrix}
\end{equation}
where $Z_2(\nu)=\Delta_p-\nu+\mathrm{j}\eta_2$ and
$Z_k(\nu)=D_k-\nu+\mathrm{j}\eta_r$, $k=3,4,5$.
The corresponding source vectors are
\begin{equation}
\begin{aligned}
\mathbf b^{(0)}
&=
\left[
\Omega_p/2,0,0,0
\right]^{\rm T},\\
\mathbf b^{(+)}
&=
\left[
0,0,\Omega_s\rho_{31}^{(0)}/2,0
\right]^{\rm T},\\
\mathbf b^{(-)}
&=
\left[
0,\Omega_s\rho_{41}^{(0)}/2,0,0
\right]^{\rm T}.
\end{aligned}
\label{eq:A_sources}
\end{equation}
The signal phase $\theta_s$ is already incorporated into
the Floquet harmonics.

Since the three Floquet components share the same matrix
structure, their solutions can be obtained through a common
elimination procedure. For a fixed $q$, let $x_k=\rho_{k1}^{(q)}$,
$k=2,3,4,5$, and write
$\mathbf b^{(q)}=[b_2,b_3,b_4,b_5]^{\rm T}$.
The subscripts follow the atomic-state indices.
In the following elimination, the superscript $q$ is omitted
and all response functions are evaluated at $\nu_q$. Specifically, eliminating the third
and fourth entries of $\mathbf x^{(q)}$, we have
\begin{equation}
x_2=
\frac{1}{F}
\left[
\Xi b_2+
\frac{\Omega_c}{2}
\left(
b_3+\frac{\Omega_Lb_4}{2Z_4}
+\frac{\Omega_Ab_5}{2Z_5}
\right)
\right],
\label{eq:A_optical_solution}
\end{equation}
where
\begin{equation}
\begin{aligned}
\Xi(\nu)
&=
Z_3(\nu)
-\frac{\Omega_L^2}{4Z_4(\nu)}
-\frac{\Omega_A^2}{4Z_5(\nu)},
\end{aligned}
\label{eq:A_Xi_F}
\end{equation}
and 
\begin{equation}
    F(\nu)
=
Z_2(\nu)\Xi(\nu)-\frac{\Omega_c^2}{4}.
\end{equation}

Substituting
$\mathbf b^{(0)}$ into~\eqref{eq:A_optical_solution}, we obtain
\begin{equation}
\begin{aligned}
\rho_{21}^{(0)}
&=\frac{\Omega_p\Xi(0)}{2F(0)},
\rho_{31}^{(0)}
=\frac{\Omega_p\Omega_c}{4F(0)},
\rho_{41}^{(0)}
=\frac{\Omega_p\Omega_c\Omega_L}
{8Z_4(0)F(0)}.
\end{aligned}
\label{eq:A_steady_solution}
\end{equation}

Similarly, substituting $\mathbf b^{(+)}$ and
$\mathbf b^{(-)}$ into~\eqref{eq:A_optical_solution}
and using~\eqref{eq:A_steady_solution}, we have
\begin{equation}
\begin{aligned}
\rho_{21}^{(+)}
&=
\frac{\Omega_s\Omega_p\Omega_c^2\Omega_L}
{32F(0)Z_4(\delta_s)F(\delta_s)},\\
\rho_{21}^{(-)}
&=
\frac{\Omega_s\Omega_p\Omega_c^2\Omega_L}
{32F(0)Z_4(0)F(-\delta_s)}.
\end{aligned}
\label{eq:A_sideband_solution}
\end{equation}
Therefore, we complete the proof.

\bibliographystyle{IEEEtran}
\bibliography{myref}

\end{document}